\documentclass[12pt,journal,draftclsnofoot, onecolumn, letterpape]{IEEEtran} %

\usepackage{amsfonts}

\usepackage[dvips]{graphicx}
\usepackage{amsmath}
\usepackage{amssymb}
\usepackage{multirow}
\usepackage{longtable}
\usepackage[tight,footnotesize, hang,raggedright, nooneline]{subfigure}
\usepackage{float}
\usepackage{afterpage}
\usepackage{multirow}
\usepackage[nospace]{cite}
\usepackage{subeqnarray}
\usepackage{algorithm}
\usepackage{algpseudocode}
\usepackage[utf8]{inputenc}
\usepackage{mdwlist}
\usepackage{tabularx}
\usepackage{color}
\usepackage{graphicx}
\usepackage{epstopdf}
\usepackage{diagbox}

\newtheorem{theo}{Theorem}
\newtheorem{prop}{Proposition}

\long\def\symbolfootnote[#1]#2{\begingroup
\def\thefootnote{\fnsymbol{footnote}}
\footnote[#1]{#2}\endgroup}
\IEEEoverridecommandlockouts
\DeclareMathSizes{12}{10}{8}{5}
\begin{document}
%
\title{
Throughput Analysis and Energy Efficiency Optimization for Standalone LTE-U Networks with Randomly Delayed CSI
}
\author{Hangguan Shan, Taojie Qin, Guanding Yu, Lin X. Cai, Yu Cheng, and Tony Q. S. Quek
}
\maketitle

\vspace{-1.3cm}
\begin{abstract}
To coexist with Wi-Fi friendly, a standalone long-term evolution network over unlicensed spectrum (LTE-U) under listen-before-talk (LBT) mechanism can only access channel in a random and intermittent way, which results in random and time-variant delay in both data and signaling transmissions. In this work, we explore the impact of randomly delayed channel state information (CSI) on the performance of a standalone LTE-U network by analyzing its downlink throughput and users’ energy efficiency (EE) under different CSI feedback schemes and frequency-domain schedulers. Further, aiming at maximizing users’ EE of a standalone LTE-U network while guaranteeing fair coexistence with Wi-Fi, joint optimization on the medium access control (MAC) protocol and  CSI feedback scheme is studied. Due to the non-convex and non-concave characteristic of the formulated optimization problems, decomposition-based low-complexity yet efficient algorithms are proposed. Simulation results not only verify the analysis and the effectiveness of the proposed algorithms but also show diverse impacts of CSI feedback scheme, frequency-domain scheduler, and traffic load of Wi-Fi on the standalone LTE-U network performance.

\end{abstract}

\vspace{-0.25cm}

\section{Introduction}

To relieve wireless radio resource scarcity and meet the unprecedented growth in wireless data, 
deploying standalone long-term evolution networks over unlicensed spectrum (LTE-U) is one of the most cost effective networking approaches \cite{3GPP_NR, Multefire}.
%
Different from licensed assisted access (LAA), a standalone LTE-U solely operates over unlicensed spectrum, i.e., both data and signaling are transmitted over unlicensed band \cite{Huang_Network18, Yuan2018TVT}. Without assistance of licensed spectrum, resource allocation in a standalone LTE-U network faces great challenges.

In the unlicensed spectrum, to be a globally good neighbor of the incumbent radio access technologies (primarily Wi-Fi networks), standalone LTE-U technology should follow harmonized standards specified in different regions including that of European Telecommunications Standards Institute (ETSI) \cite{EtsiBroadband}.
%
%
%
For coexistence purpose, the basic functionalities it needs to comply with include listen-before-talk (LBT)-based channel access and maximum channel occupancy time (MCOT) constrained on the duration of each channel access.
For fair unlicensed spectrum access, according to 3GPP Release 14, MCOT is 10ms \cite{3GPP36_213}.
%
On the other hand, to take advantage of channel-aware transmissions exploited by LTE systems, a base station (BS) needs to collect channel state information (CSI) from users. To achieve this, the BS needs to select a set of users and sends reference signals to them; then the selected users access the channel to report the measured CSI. As  LTE-U uses a contented based channel access protocol, the BS and users usually wait for a random time before accessing the channel, which results in relatively long and randomly deplayed CSI feedback.


Most research works in LTE-U mainly focus on medium access control (MAC) protocol design (e.g., \cite{Zhang2015LTE, Chen2016Coexistence, Khairy_Globecom17, Chen2017Embedding}) and resource allocation (e.g., \cite{He2016Proportional, Zhang_JSAC16, Zhen_VTC2018, Liu2014Small, Chen_TWC16}). In these works performance metrics are analyzed or evaluated by either assuming fixed transmission rate \cite{Khairy_Globecom17} or perfect CSI without feedback delay \cite{Liu2014Small, He2016Proportional, Zhang_JSAC16, Chen_TWC16, Chen2017Embedding, Zhen_VTC2018}. To the best of our knowledge, very few works take account of the impact of randomly delayed CSI feedback. It is found in \cite{Guharoy2013Joint, Ananya2015Performance} that, though feedback delays in traditional LTE networks are usually not random, the delay might result in high outage probability and degrade the network throughput.
As such, to understand the impact of randomly delayed CSI feedback on the performance of a standalone LTE-U network further research is indispensable.

In this paper, 
we study the impact of randomly delayed CSI feedback on the performance of a standalone LTE-U network by analyzing its downlink (DL)  throughput and users' energy efficiency (EE) under different frequency-domain schedulers and CSI feedback schemes.
Specifically, four frequency-domain schedulers, round robin scheduler, greedy scheduler, proportional fairness (PF) scheduler, and random scheduler, and two CSI feedback schemes,
threshold-based feedback scheme and best-$m$ feedback scheme, are studied.
A comprehensive analytical framework integrated with the aforesaid schedulers and feedback schemes is proposed to unearth the two performance metrics of interest of a standalone LTE-U network in both independent and identically distributed (i.i.d.) and non-i.i.d. user scenarios, respectively.
The main contributions and significance of this paper are three-fold.

Firstly, we study how to model the randomly delayed CSI feedback for a standalone LTE-U network. 
Based on the frame structure, duplex mode, and utilizing the probability mass function (pmf) of the duration of channel contention obtained via the Lattice-Poisson algorithm as studied in our previous work \cite{Li2017Modelling}, we derive the distribution of the CSI feedback delay 
under the condition that both DL and uplink (UL) of the network adopt the same LBT-based channel access scheme.

Secondly, we shed some light on how different frequency-domain schedulers and/or CSI feedback schemes under randomly delayed CSI feedback affect the network throughput and users' EE of a standalone LTE-U network. It is unveiled  that to boost the network throughput different schedulers can prefer opposite setting of a feedback scheme.
For channel-aware transmissions, the impact of imperfect CSI due to feedback delay can deteriorate the network throughput, as Wi-Fi load increases. It is possible that a feedback scheme's setting benefits network throughput but degrades users' EE. 
Further, the greedy scheduler does not always work best if the feedback delay is large.

Last but not least, we study cross-layer parameter optimization of the MAC protocol and CSI feedback scheme to maximize users' EE of the standalone LTE-U network while guaranteeing fair coexistence with Wi-Fi. 
However, as the objective functions of the formulated problems are non-convex and non-concave, we propose decomposition-based low-complexity yet efficient algorithms to solve the optimization problems (OPs).
Simulation results not only show the effectiveness of the proposed algorithms, but also find that, to improve users' EE of the standalone LTE-U network while protecting Wi-Fi of high traffic load, the standalone LTE-U network should interact with Wi-Fi by reducing channel access chances but increase its users' feedback opportunities simultaneously.

The remainder of this paper is organized as follows. Related work is presented in Section \ref{sect:realated work}. The system model is described in Section \ref{sect:system model}. Network throughput and users' EE in various scenarios are analyzed in Sections \ref{sect: throughput analysis} and \ref{sect: EE analysis}, respectively. In  Section \ref{sect: EE optimization},
EE optimization is studied.
Performance evaluation is provided in Section \ref{sect: simulation results}, followed by the conclusions in Section \ref{sect: conclusion}.

\section{Related Work}
\label{sect:realated work}

To understand the coexistence performance of the LTE-U networks analytically, stochastic geometry based approaches are exploited in the literature \cite{Li_TWC16, Wang_JSAC17, Ajami_TWC17}. Focusing on a scenario of DL traffic only, stochastic geometry based framework is proposed in \cite{Li_TWC16} to compare three LTE coexistence mechanisms, LTE with continuous transmission, LTE with discontinuous transmission, and LTE with LBT and random backoff, from the perspectives of medium access probability, signal-to-interference-plus-noise ratio coverage probability, density of successful transmission, and rate coverage probability. Further, assuming synchronized and slotted carrier sense multiple access with collision avoidance (CSMA/CA) in Wi-Fi and carrier censing adaptive transmission (CSAT) in LTE-U, the analysis in \cite{Wang_JSAC17} reveals that the density of Wi-Fi APs has no effect on the asymptotic DL spatial throughput of LTE-U while a large LTE-U BS density enlarges its own DL spatial throughput but diminishes that of Wi-Fi. The study in \cite{Ajami_TWC17} extends the analysis of the aforesaid two works into a scenario of simultaneous UL and DL Wi-Fi transmissions using the IEEE 802.11ax standard with both single user  and multi-user  operation modes.

To enable better coexistence of LTE-U and Wi-Fi, in the literature multiple efficient approaches are proposed (e.g., \cite{Wang_JSAC17, Zhang_JSAC16, He2016Proportional, Yuan_JSAC17, Yin_TWC18, Li_TVT17, Ko_TVT18}).
Based on the analysis of asymptotic spatial throughput, proposed in \cite{Wang_JSAC17} is to optimize the retention probability of LTE-U nodes thereby maximizing the minimum weighted spatial throughput of LTE-U and Wi-Fi.
In \cite{Zhang_JSAC16}, a channel access model for LTE-U and Wi-Fi  in unlicensed bands is established by Markov chain-based analysis, whereby fairness-based LAA channel access approach is designed and the optimal contention window (CW) size for the LTE-U networks is obtained. 
%
To achieve throughput-oriented PF between LTE-U and Wi-Fi, proposed in \cite{He2016Proportional} is a cross-layer optimization framework which jointly optimizes the MAC layer protocol parameters as well as channel and power allocation of an LTE-U network. 
It is found in \cite{Yuan_JSAC17} that there is a significant benefit in deploying adaptive spectrum partitioning between Wi-Fi and LTE-U for improving human satisfaction.
In \cite{Yin_TWC18}, by considering that the LTE-U BSs can decode control packets from Wi-Fi thus knowing the CSI to the Wi-Fi users and AP, multi-antenna transmit beamforming is studied to enable spatial reuse for coexisting the two networks. 
In \cite{Li_TVT17}, a distributed algorithm is proposed to adaptively change the energy detection threshold for LAA, 
so that the system can encourage more concurrent transmissions while avoiding collisions.
Different from the aforesaid works, in \cite{Ko_TVT18} to maximize LTE-U's throughput while maintaining fairness between LTE-U and Wi-Fi, channel selection is jointly optimized with frame scheduling in LTE-U.

Though the aforementioned theoretical analyses and novel mechanism design offer insights and guidelines to enhance or optimize the coexistence of LTE-U and Wi-Fi, their solutions either ignore the impact of randomly delayed CSI feedback or use perfect channel condition. So, one goal of this paper is to establish an analytical framework to unearth the hidden impact and thus facilitating better coexistence of the two networks.
It is noteworthy that there exist some previous works developing comprehensive analysis for orthogonal frequency-division multiple access (OFDMA) system  in the presence of feedback delays. For example, assuming i.i.d. subchannel gain of each user, joint evaluation of channel feedback scheme, rate adaption, and scheduling is performed in \cite{Guharoy2013Joint} to analyze the impact of feedback delays on the throughput and outage probability of the system.
\cite{Ananya2015Performance} extends the work by further considering a case in which the subchannel gains are uniformly correlated.
As users may have different channel statistics, in \cite{Huang2013Performance} heterogeneous feedback design with different subband sizes is proposed and analyzed. 
However, the results of these works cannot be directly applied to the standalone LTE-U networks, as the randomly delayed CSI feedback due to the MAC layer interaction between the coexisting networks is excluded in their modeling.

%
%

Similar to Wi-Fi, when LBT is implemented in the LTE-U networks, both BSs and UEs spend more energy to sense or listen to the channel, generating EE issue \cite{Li2017Modelling, Maheshwari_TMC19, Luo_CL15}. 
In \cite{Li2017Modelling}, the authors present a MAC-layer analytical framework to understand the synchronisation performance and the involved energy consumption of initial and maintaining synchronisation in a standalone LTE-U network.
Similar to the configuring discontinuous reception (DRX) mechanism in LTE, \cite{Maheshwari_TMC19} proposes LAA-DRX mechanism to reduce UE's energy consumption.
To achieve a beneficial balance between energy consumption and throughput gain, proposed in \cite{Luo_CL15} is a dynamic carrier component on/off scheduling scheme.
Different from the aforesaid works, \cite{Chen2016Energy} studies the physical-layer cause of reducing EE when using unlicensed spectrum, the larger channel attenuation as compared with the licensed spectrum.
However, the impact of randomly delayed CSI feedback is ignored when modeling the data rate and users' EE of the LTE-U networks in the aforementioned  works. So, the other goal of this paper
is to understand this issue and design cross-layer approach to improve users' EE in the networks.

\section{System Model}
\label{sect:system model}

\begin{figure}
  \centering
  \includegraphics[width=0.45\linewidth]{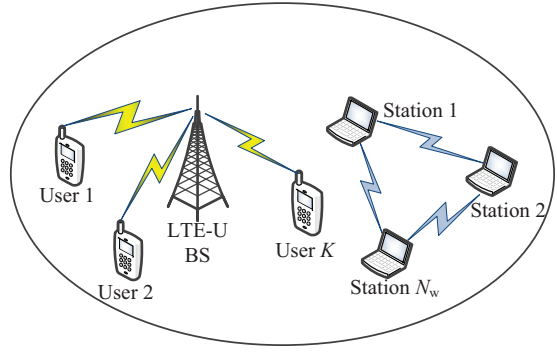}
    \vspace{-0.35cm}
  \caption{System model.}
  \label{fig:systemmodel}
  \vspace{-0.2cm}
\end{figure}

We consider a standalone LTE-U network and a Wi-Fi network sharing an unlicensed channel with bandwidth $B$. As shown in Fig. \ref{fig:systemmodel}, the Wi-Fi network is composed of $N_{\rm w}$ Wi-Fi stations (STAs) and the standalone LTE-U network consists of a BS and $K$ single-antenna users. All nodes in the network are assumed to have saturated traffic.
The standalone LTE-U network divides the channel into $S$ orthogonal subchannels each with bandwidth $B/S$.
Let $s$ (resp. $k$) and $\cal S$ (resp. $\cal K$) denote the subchannel (resp. user) index and subchannel (resp. user) set, respectively.
Similar to an LTE network, for DL channel-aware transmissions the BS first sends a UL grant signal to inform users of reporting the CSI.
The users received the grant signal estimate the channel and report the CSI according to the adopted CSI feedback scheme when they access channel.
After receiving the CSI, the scheduler at the BS can allocate resources for data transmissions. The detailed frame structure, channel contention scheme, channel and rate adaption model, frequency-domain scheduler, and CSI feedback scheme are discussed as follows.



\begin{figure}
  \centering
  \includegraphics[width=0.9\linewidth]{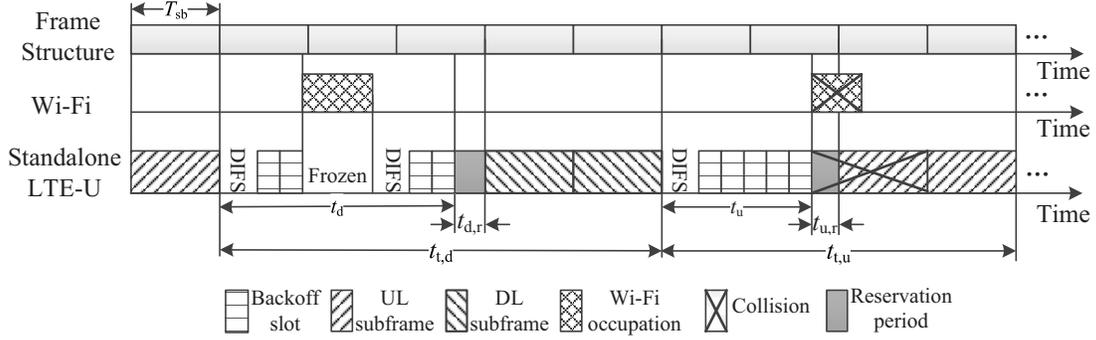}
  \vspace{-0.35cm}
  \caption{Frame structure and LBT channel contention scheme.}
  \label{fig:framestructure}
  \vspace{-0.75cm}
\end{figure}

\subsection{Frame Structure and Channel Contention Scheme}
Consider that the standalone LTE-U network works in a time-division duplex (TDD) mode with a frame structure as shown in Fig. \ref{fig:framestructure}. So, both the DL and UL transmissions occur in the same unlicensed channel and are time slotted. Here, the slot of duration $T_{\rm {sb}}$ is referred to as the subframe used in an LTE system. However, to coexist with Wi-Fi friendly, the LTE-U network cannot always occupy channel but use LBT or on/off-based  channel contention schemes \cite{Zhang2015LTE}. Here, we assume that the same LBT channel contention scheme, based on a fixed-CW size backoff mechanism (i.e., category 3 defined in \cite{LAA2015Study}), is employed for both DL and UL transmissions in the LTE-U network.

Specifically, to initiate a 
DL data transmission, 
after the last subframe of the previous UL data transmission, the BS waits for a distributed inter-frame spacing (DIFS) and then chooses a backoff counter randomly distributed between zero and the value of $Z-1$ of the CW.
The backoff counter decreases by 1 in every subsequent backoff slot,\footnote{A backoff slot is the short slot with the same interval of fixed duration specified by the IEEE 802.11 standard \cite{IEEE2010IEEE}.} as long as the channel is detected idle in that backoff slot.
Otherwise, the BS freezes its backoff counter.
The count resumes when the channel is idle for a DIFS interval again. When the backoff counter reaches zero, the BS finishes channel contention and starts data transmissions.
Let $t_{\rm d}$ denote the duration of this \emph{DL channel contention} (see Fig. \ref{fig:framestructure}).
Notice that the BS can complete the contention procedure at any time.
For consistency with the frame structure (i.e., starting transmission at the boundary of a subframe), if any, to avoid Wi-Fi seizing channel the BS sends a reservation signal before the arrival of the next subframe (see Fig. \ref{fig:framestructure}).
The duration of this period is denoted as $t_{\rm d, r}$ and called \emph{DL reservation period}.
Then, the overall time that the BS consumes before transmitting data (named \emph{DL pre-transmission period}) is $t_{\rm d, p} = t_{\rm d} + t_{{\rm d, r}}= \left\lceil {t_{\rm d}/T_{\rm sb}} \right\rceil T_{\rm sb}$, where $\left\lceil  \cdot  \right\rceil $ is the ceiling function.
When transmitting data (referred to as \emph{effective DL transmission}), the BS can transmit data up to $N_{\rm d}$ subframes.
Therefore, the total duration of a 
DL data transmission can be represented as $t_{\rm t,d} =t_{\rm d,p}+ N_{\rm d}T_{\rm sb}$.

For the UL data transmissions, the channel contention procedure is similar to the aforesaid DL case. The main difference is that, although the users granted by the BS will fulfill channel contention, the backoff counter is selected by the BS (again between zero and $Z-1$) and shared by all those users together.
The BS can send the information in one subframe of the last DL transmission.
So, there can be no intra-collisions among users in the standalone LTE-U network.
Further, in the UL case the users rather than the BS will send the reservation signal.
Similar to the DL case, we let 
$t_{\rm u}$, $t_{\rm u, r}$, $t_{\rm u, p}$, $N_{\rm u}$, and $t_{\rm t,u}$  denote 
the duration of a \emph{UL channel contention}, the duration of a \emph{UL reservation period},
the overall time for users contending channel and sending reservation signals (referred to as \emph{UL pre-transmission period}), the number of subframes allowing users to transmit data (referred to as \emph{effective UL transmission}), and the total duration of a 
UL data transmission, respectively.
To simplify analysis, in this work we assume $N_{\rm d} = N_{\rm u} = N_{\rm sb}$.


\subsection{Channel and Rate Adaption Model}

Let $G_{k,s}$ denote the channel power gain user $k$ estimates for subchannel $s$ from the BS's one DL  transmission. To acquire the latest CSI for rate and subchannel assignment, in the standalone LTE-U network of interest, the users should estimate the channel in the last subframe of the BS's DL transmission.
Let $G_{k,s,\alpha}^{\rm d}$ denote the actual subchannel gain when the BS transmits data for user $k$ according to $G_{k,s}$ in subframe $\alpha$ ($\in \{1,2,...,N_{\rm sb}\}$) of the next DL transmission, and $\tau _\alpha$ be the involved feedback delay 
between channel estimation and data transmission.
We consider a wide sense stationary Rayleigh fading process. Therefore, $G_{k,s}$ and $G_{k,s,\alpha}^{\rm d}$ are correlated exponential random variables with the same mean $\Omega _k$.
Their joint probability density function (pdf) given channel estimation delay $\tau _\alpha$ is given by \cite{Simon2000Digital}
\begin{align} \label{eq: joint pdf}
{f_{{G_{k,s}},{G_{k,s,\alpha}^{\rm d}} |\tau_\alpha}}( {x,y|z} ) = \frac{\exp \left( { - \frac{{x + y}}{{{\Omega _k}( {1 - \rho (z) } )}}} \right)}{{\Omega _k^2( {1 - \rho (z) } )}}{I_0}\left( {\frac{{2\sqrt{\rho (z)}  \sqrt {xy} }}{{{\Omega _k}( {1 - \rho (z) } )}}} \right), x,y,z \ge 0
\end{align}
where 
$I_0\left( \cdot \right)$ is the 0-th-order modified Bessel function of the first kind, and $\rho (z)  = J_0^2(2\pi {\phi _{\rm d}}z)$ is the correlation coefficient, with $\phi _{\rm d}$ being the maximum Doppler spread and $J_0\left( \cdot \right)$ being the 0-th-order Bessel function of the first kind.

When transmitting data, the BS adapts the data rate of each channel according to the CSI of the user selected by the frequency-domain scheduler (to be discussed in Section \ref{subsect:scheduler}).
The data rate associated with each CSI feedback is determined by discretizing the range of channel fading levels. Specifically, we divide the range of subchannel gain into $N+1$ fading regions $R_n = [L_{n}, L_{n+1}), n = 0, 1, ..., N$, where $L_0 = 0$ and $L_{N+1} = \infty$. Then, the BS transmits data with rate $r_n$ when $G_{k,s} \in R_n$; accordingly, to receive successfully we need $G_{k,s,\alpha}^{\rm d} \ge L_n$.

\subsection{Frequency-Domain Schedulers}
\label{subsect:scheduler}

Four different frequency-domain schedulers, round robin, greedy scheduler, PF scheduler, and random scheduler, are studied in this work.
For each subchannel, the round robin scheduler serves users one by one in order; 
the greedy scheduler, the PF scheduler, and the random scheduler select the user, among those fed back subchannel gain for this subchannel, with the maximum subchannel gain, the maximum normalized subchannel gain (defined as the ratio of the subchannel gain to its mean \cite{Choi2007Cell}), and an equal probability, 
respectively.
Here, for the random scheduler, as only the users that fed back CSI for the subchannel can be selected, it is different from the round robin scheduler. 
In addition, we will compare it with the greedy scheduler in terms of the tolerance of feedback delay, which could be long if the standalone LTE-U network needs to coexist (content channel) with a high-load Wi-Fi network.

\subsection{Subchannel Gain Feedback Schemes}
\label{sub: Sub-gain f-schemes}

When the standalone LTE-U network occupies the channel, Wi-Fi might access channel simultaneously, resulting in a collision as shown in Fig. \ref{fig:framestructure}.
As the duration of a collided Wi-Fi transmission is usually shorter than a subframe used in an LTE system \cite{Ong2011IEEE}, only the first subframe of the DL or UL effective transmission might occur collisions.
Thus, to prevent grant signals and CSI report from colliding with Wi-Fi transmissions,  they should not be transmitted in the first
subframe of the DL and UL effective transmissions, respectively.
Besides, to facilitate analysis, similar to \cite{Guharoy2013Joint, Ananya2015Performance} we assume that both channel estimation and CSI feedback are error-free.
Two subchannel gain feedback schemes, threshold-based feedback and best-$m$ feedback, are studied.
%
For threshold-based feedback scheme, a user reports the subchannel gain of a subchannel only if the subchannel gain (resp. normalized subchannel gain) exceeds a threshold $\lambda$ for the greedy, round robin, or random scheduler (resp. PF scheduler).
For the best-m feedback scheme, a user feeds back its $m$ highest subchannel gains and indices among all subchannels.

\section{Throughput Analysis}
\label{sect: throughput analysis}

In this section, we develop an analytical model to study the impact of randomly delayed CSI on the throughput of the standalone LTE-U network. 
Firstly, we derive 
the distribution of the feedback delay and the probability of subframe collision, based on our previous work \cite{Li2017Modelling} which uses the Lattice-Poisson algorithm to calculate the pmf of the duration of channel contention of a standalone LTE-U network.
Then, we characterize the network throughput under different user scenarios and scheduler-feedback scheme combinations.



\subsection{Feedback Delay Distribution}
\label{sect: feedback delay distribution}

The feedback delay $\tau_{\alpha}$ consists of four parts: the UL pre-transmission period $t_{\rm u,p}$, the effective UL transmission period $N_{\rm sb}T_{\rm sb}$, the DL pre-transmission period $t_{\rm d,p}$, and the duration of the first $(\alpha-1)$ subframes in the  effective DL transmission $(\alpha-1)T_{\rm sb}$, i.e.,
%
\begin{equation}\label{eq:taualfa}
    \tau_{\alpha} = t_{\rm u,p} + N_{\rm sb} T_{\rm sb} + t_{\rm d,p}  + (\alpha-1)T_{\rm sb}.
\end{equation}
Because both random variables (RVs) $t_{\rm u,p}$ and $t_{\rm d,p}$ are not less than the length of a subframe (due to the existence of DIFS after each channel occupancy), we further have
\begin{equation}
\tau_{\alpha} \ge T_{\rm sb} + N_{\rm sb} T_{\rm sb} + T_{\rm sb} + (\alpha-1)T_{\rm sb} = N_{\rm min}T_{\rm sb}
\end{equation}
where $N_{\rm min}= N_{\rm sb}+ \alpha +1$.
So, the distribution of feedback delay $P\{\tau _\alpha =a T_{\rm sb}\}$, $a \in [N_{\rm min}, \infty) \cap  {\mathbb Z}^+$, determined by the distributions of  $t_{\rm u,p}$ and $t_{\rm d,p}$, $P\{t_{\rm u,p} = b_{\rm u} T_{\rm sb}\}$ and $P\{t_{\rm d,p} = b_{\rm d} T_{\rm sb}\}$, $b_{\rm u}, b_{\rm d} \in {\mathbb Z}^+$, can be derived as
\begin{equation} \label{eq: pro of tau alpha}
\begin{split}
P\{ \tau _\alpha = a T_{\rm sb} \} &= \sum\limits_{b_{\rm u} \in {\mathbb Z}^+ } {P\{{t_{\rm u,p}}=b_{\rm u} T_{\rm sb}, {t_{\rm d,p} = a T_{\rm sb} - { t_{\rm u,p}} - (N_{\rm sb} +\alpha -1){T_{\rm sb}}} \}}\\ 
&\mathop  = \sum\limits_{b_{\rm u} \in {\mathbb Z}^+ } {P\{{t_{\rm u,p}}=b_{\rm u} T_{\rm sb}\}P\{ {t_{\rm d,p} = a T_{\rm sb} - b_{\rm u} T_{\rm sb} - (N_{\rm sb} +\alpha -1){T_{\rm sb}}} \}}\\
&\mathop  = \sum\limits_{b_{\rm u} \in {\mathbb Z}^+ } {P\{{t_{\rm u,p}}=b_{\rm u} T_{\rm sb}\}P\{ { t_{\rm d,p} = ( a - b_{\rm u}  - {{N_{\rm min}} + 2} ){T_{\rm sb}}} \}}\\
\end{split}
\end{equation}%
where the second equality holds since RVs $t_{\rm u,p}$ and $t_{\rm d,p}$ are independent (as their components $t_{\rm u}$ and $t_{\rm d}$ are i.i.d. RVs due to random selection of the backoff counters within the same CW).
Without differentiation, let $p(t)$  denote the pmf of $t_{\rm u}$ and $t_{\rm d}$. Given saturated network traffic, it mainly depends on the node size of  Wi-Fi and the MAC protocol parameters of both Wi-Fi and LTE-U networks (e.g., the CW size $Z$ of the LTE-U network),
but can be obtained by probability generating function-based Lattice-Poisson algorithm proposed in \cite{ Li2017Modelling, sakurai2007mac}.
%
%
%
%
Therefore, the distributions of $t_{\rm u,p}$ and $t_{\rm d,p}$ can be calculated as
\begin{align} \label{eq: pro of pip}
P\{{t_{i, \rm p}} = b_{i} T_{\rm sb}\} = \sum\limits_{t \in\left((b_i-1)T_{\rm sb}, b_i T_{\rm sb}\right]} p(t), ~ i \in \{\rm u, d\}.
\end{align}

Substituting (\ref{eq:taualfa}) and (\ref{eq: pro of pip}) in (\ref{eq: pro of tau alpha}) yields
\begin{equation} \label{eq: pro of tau alpha2}
P\{ \tau _\alpha = a T_{\rm sb} \} = \sum\limits_{ b_{\rm u} = 1 }^{a + 1 - {N_{\rm min}}} \sum \limits_{t \in\left((b_{\rm u}-1)T_{\rm sb}, b_{\rm u} T_{\rm sb}\right]} p(t) \sum \limits_{t \in \left((a - b_{\rm u}-N_{\rm min} + 1)T_{\rm sb}, (a - b_{\rm u}-N_{\rm min} + 2)T_{\rm sb}\right]} p(t).
%
\end{equation}

\subsection{Subframe Collision Probability}
\label{subsect: collision pro}

Though the reservation signal sent in the DL or UL reservation period helps reduce collisions between LTE-U and Wi-Fi, collisions may occur in the effective data transmission period.
But, as the duration of a collided Wi-Fi transmission (denoted by $T^{\rm w}_{\rm c}$) is shorter than a subframe's length $T_{\rm sb}$, only the first subframe in the DL or UL effective data transmission might occur collisions.
Depending on the relationship between the length of the reservation period ($t_{i, {\rm r}}$, $i \in \{\rm d, u\}$) and the collided duration $T^{\rm w}_{\rm c}$, one knows that: if $t_{i, \rm r} \geq T^{\rm w}_{\rm c}$, the first subframe will not be collided; otherwise, the first subframe might have a collision depending on whether or not Wi-Fi STAs transmit at the beginning of the reservation period.
%
%
As $t_{\rm d, r}$ and $t_{\rm u, r}$ are respectively decided by $t_{\rm d}$ and $t_{\rm u}$ and the latter two are i.i.d. RVs, the collision probabilities of the DL and UL cases are equal.
Without distinguishing DL and UL, let $p_{{\rm c},1}$ and $p_{\rm L}$ denote the collision probability of the first subframe of the effective data transmission and the collision probability when the BS or the users send reservation signals in a reservation period, respectively.
Therefore, we have
\begin{align} \label{eq: collision pro of first subframe}
\begin{split}
{p_{{\rm c},1}} &= {p_{\rm L}} P\{ 0 \leq t_{ i,\rm r} < T_{\rm c}^{\rm w}\}
= p_{\rm L} P\{0 \leq t_{i,\rm p}-t_{i} < T_{\rm c}^{\rm w}\} \\
&= p_{\rm L} \sum\limits_{b_i \in {\mathbb Z^ + }} P\{ b_{i} T_{\rm sb} - T_{\rm c}^{\rm w} \leq t_{i}   < b_{i} T_{\rm sb} \} 
= {p_{\rm L}} \sum\limits_{ b_{i} \in {\mathbb Z^ + }} \sum\limits_{t \in \left(b_i T_{\rm sb} - T_{\rm c}^{\rm w}, b_i T_{\rm sb} \right]} p(t)
\end{split}
\end{align}
where $i \in \{\rm d, u\}$ and $p_{\rm L}$ can be derived numerically according to Appendix \ref{app:deduce tau pL}.
Then, for any subframe, subframe collision probability can be summarized as ${p_{{\rm c}, \alpha}} =p_{\rm c,1}$, if $\alpha  = 1$, otherwise if $\alpha  = 2,3,...,{N_{\rm sb}}$, it is 0.
%

\subsection{Network Throughput}
As the standalone LTE-U network works in a TDD mode, its DL network throughput accumulated by all subchannels and subframes 
can be defined as follows
\begin{equation} \label{eq: network throughput}
\begin{split}
\eta _{\rm{L}}^{\gamma ,\vartheta } 
& = \frac{{{T_{\rm sb}}B}}{S({\mathbb E}(t_{\rm t,d})+ {\mathbb E}(t_{\rm t, u}))}\sum\limits_{\alpha  = 1}^{{N_{\rm sb}}} {\sum\limits_{s = 1}^S {\mathbb E} ({\eta _{s,\alpha }^{\gamma ,\vartheta }} ) }
= \frac{{{T_{\rm sb}}B}}{{\mathbb E}(t_{\rm t,d})+{\mathbb E}(t_{\rm t,u})}\sum\limits_{\alpha  = 1}^{{N_{\rm sb}}} {{\mathbb E}({\eta _{s,\alpha }^{\gamma ,\vartheta }}) }\\
& = \frac{{{T_{\rm sb}}B}}{{\mathbb E}(t_{\rm t,d})+{\mathbb E}(t_{\rm t,u})}\sum\limits_{\alpha  = 1}^{N_{\rm sb}} {\bar p_{{\rm c},\alpha }}\sum\limits_{a \in [N_{\rm min}, \infty) \cap  {\mathbb Z}^+} {\mathbb E} (\eta _{s,\alpha }^{\gamma ,\vartheta}| \tau _\alpha = a T_{\rm sb}) P\{ \tau _\alpha = a T_{\rm sb} \}
\end{split}
\end{equation} %
where
$\gamma \in \left\{\rm RR, G, PF,R \right\}$ represents the adopted frequency-domain scheduler, with RR, G, PF, and R denoting the round robin scheduler, the greedy scheduler, the PF scheduler, and the random scheduler, respectively;
$\vartheta \in \left\{\rm TH, BM\right\}$ represents subchannel gain feedback scheme, with TH and BM representing threshold-based feedback  and best-$m$ feedback, respectively;
${\mathbb E}(t_{{\rm t},i})$, $i \in \{\rm d, u\}$, denotes the mean time duration of a DL or UL data transmission, satisfying
\begin{align} \label{eq: TD TU}
{\mathbb E}(t_{{\rm t}, i})=\sum\limits_{ b_{i} \in {{\mathbb Z}^ + }} \sum\limits_{t \in \left((b_i-1)T_{\rm sb}, b_i T_{\rm sb}\right]} b_i T_{\rm sb} p(t)  + {N_{\rm sb}}{T_{\rm sb}},~i \in \{\rm d,u\};
\end{align}
${\mathbb E}(\eta _{s, \alpha}^{\gamma, \vartheta})$ calculates the mean of transmission rate $\eta _{s, \alpha}^{\gamma, \vartheta}$ in the $s$-th subchannel at the $\alpha$-th subframe during a DL data transmission when the standalone LTE-U network adopts $\gamma$ scheduler and $\vartheta$ feedback scheme;
${\bar p_{{\rm c},\alpha }} = 1-{p_{{\rm c},\alpha }}$ gives the probability that subfame $\alpha$ is not in a collision.
%

In (\ref{eq: network throughput}), the second equality holds because the subchannel gains are statistically identical for each user, 
i.e., $ {\mathbb E}(\eta _{s,\alpha }^{\gamma ,\vartheta }) = {\mathbb E} (\eta _{s', \alpha }^{\gamma ,\vartheta }), s, s' \in {\cal S}, \alpha = 1, 2, ..., N_{\rm sb}$; and the last equality holds as we consider only non-collided subframes contribute throughput and the mean transmission rate should be calculated by averaging on all possible feedback delays.

Finally, to obtain the DL  throughput of the LTE-U network $\eta _{\rm{L}}^{\gamma ,\vartheta }$, one needs to calculate ${\mathbb E} (\eta _{s,\alpha }^{\gamma ,\vartheta}| \tau _\alpha = a T_{\rm sb})$.
For both the i.i.d. user scenario (defined as $\Omega_k = \Omega, k\in {\mathcal K}$) and non-i.i.d. user scenario, the analytical results of  ${\mathbb E} (\eta _{s,\alpha }^{\gamma ,\vartheta}| \tau _\alpha = a T_{\rm sb})$
under scheduler-feedback scheme combinations from $\{\rm RR, G, PF\} \times \{\rm TH, BM\}$ have been studied in \cite{Guharoy2013Joint}.
So, in below we only derive the results for the random scheduler. 

\begin{theo} \label{theo:1}
%
For threshold-based feedback with the random scheduler and i.i.d. users, given feedback delay $\tau _\alpha$, the mean transmission rate in the $s$-th subchannel at the $\alpha$-th subframe is 
\begin{align} \label{eq: throughput of R TH iid}
\begin{split}
{\mathbb E} (\eta _{s,\alpha }^{\rm R,TH}| \tau _\alpha)= & \sum\limits_{\delta  = 1}^K  \binom{K}{\delta}\left( 1 - {
\rm exp} \left(  - \frac{\lambda }{\Omega} \right) \right)^{K - \delta } {\rm exp}\left(  - \frac{\lambda \left( \delta - 1 \right) }{\Omega } \right) \sum\limits_{n = \omega }^N {r_n} \\
& \times  \int_{{\rm max}\left\{ {L_n, \lambda } \right\}}^{L_{n + 1}} \int_{L_n}^\infty  \frac{\exp \left( { - \frac{{x + y}}{\Omega( {1 - \rho (\tau _\alpha) } )}} \right)}{\Omega ^2( 1 - \rho (\tau _\alpha)  )}{I_0}\left( \frac{2\sqrt{\rho (\tau _\alpha)}  \sqrt {xy} }{\Omega( {1 - \rho (\tau _\alpha)} )} \right) dy dx
\end{split}
\end{align}
where $\omega$ is the index such that $L_{\omega} \le \lambda < L_{\omega + 1}$ holds.
\end{theo}

\begin{IEEEproof}
The proof is provided in Appendix \ref{app:theo:1}.
\end{IEEEproof}

\begin{theo} \label{theo:2}
For best-$m$ feedback with the random scheduler and i.i.d. users, given feedback delay $\tau _\alpha$, the mean transmission rate in the $s$-th subchannel at the $\alpha$-th subframe is 
\begin{equation} \label{eq: throughput of R BM iid}
\begin{split}
{\mathbb E} (\eta _{s,\alpha }^{\rm R,BM}| \tau _\alpha) =& \sum\limits_{\delta  = 1}^K  {\binom{K}{\delta}{\left( \frac{m}{S} \right)^{\delta - 1 }}{\left(1-\frac{m}{S}\right)^{K-\delta}}} \sum\limits_{n = \omega }^N {{r_n}} \int_{{\rm max}\left\{ {{L_n},\lambda } \right\}}^{{L_{n + 1}}} \int_{{L_n}}^\infty    {\varpi _\Omega }\left( x \right)  \\
&  \times   \frac{\exp \left(  - \frac{{x + y}} {\Omega( 1 - \rho  (\tau _\alpha)  )} \right)} {\Omega ^2( 1 - \rho (\tau _\alpha)  )} {I_0}\left( \frac{2\sqrt{\rho (\tau _\alpha)}  \sqrt {xy} } { \Omega ( 1 - \rho (\tau _\alpha) )} \right) dy dx
\end{split}
\end{equation}
where
%
${\varpi _\Omega }( x ) = \sum\nolimits_{n = 0}^{m - 1} \binom{S-1}{n} {\rm exp}\left(  - \frac{xn}{\Omega } \right) \left( 1 - {\rm exp} \left(  - \frac{x}{\Omega} \right) \right)^{S-1-n}.$
%
\end{theo}

\begin{IEEEproof}
The proof is provided in Appendix \ref{app:theo:2}.
\end{IEEEproof}

\begin{theo} \label{theo:3}
For threshold-based feedback with the random scheduler and non-i.i.d. users, given feedback delay $\tau _\alpha$, the mean transmission rate in the $s$-th subchannel at the $\alpha$-th subframe is 
\begin{align} \label{eq: throughput of R TH niid}
\begin{split}
{\mathbb E} (\eta _{s,\alpha }^{\rm R,TH}| \tau _\alpha) = &    \sum\limits_{\delta  = 1}^K \sum\limits_{v  = 1}^{\binom{K}{\delta}}  \prod\limits_{k \in {\cal U}_{s, v} ^{\rm TH} } {{\rm exp}\left(  - \frac{\lambda }{\Omega _k} \right)} \prod\limits_{\hat k \in {\cal K}\backslash  {\cal U}_{s, v} ^{\rm TH} } \left( {1 - {\rm exp}\left(  - \frac{\lambda }{\Omega _{\hat k}} \right)} \right) \sum\limits_{n = \omega }^N {r_n}  \sum\limits_{k \in  {\cal U}_{s, v} ^{\rm TH} }  \frac{1}{\delta}\\
&   \times  {\rm exp} \left(\frac{\lambda}{\Omega_{k}}\right)  \int_{{\rm max}\left\{ {{L_n}, \lambda } \right\}}^{L_{n + 1}} \int_{L_j}^\infty  \frac{\exp \left(  - \frac{{x + y}} {{\Omega _k}( {1 - \rho (\tau _\alpha) } )} \right)} {\Omega _k^2( 1 - \rho (\tau _\alpha)  )} {I_0}\left( \frac{2\sqrt{\rho (\tau _\alpha)}  \sqrt {xy} } { {\Omega _k}( 1 - \rho (\tau _\alpha))} \right) dy dx
\end{split}
\end{align}
where $ {\cal U}_{s, v} ^{\rm TH}$ denotes the $v$-th user set containing $\delta= |{\cal U}_{s, v} ^{\rm TH}|$ users which all fed back channel gain for subchannel $s$ in the previously UL transmission, with $v = 1,2,\ldots, \binom{K}{\delta}$.
\end{theo}
%

\begin{theo} \label{theo:4}
For best-$m$ feedback with the random scheduler and non-i.i.d. users, given feedback delay $\tau _\alpha$, the mean transmission rate in the $s$-th subchannel at the $\alpha$-th subframe is 
\begin{align} \label{eq: throughput of R TH}
\begin{split}
{\mathbb E} (\eta _{s,\alpha }^{\rm R, BM}| \tau _\alpha) = &  \sum\limits_{\delta  = 1}^K \sum\limits_{v  = 1}^{\binom{K}{\delta}}  {{\left( {\frac{m}{S}} \right)}^{\delta  - 1}}{{\left( {1 - \frac{m}{S}} \right)}^{K - \delta }}  \sum\limits_{n = 1}^N   {r_n}   \sum\limits_{k \in {\cal U}_{s,v} ^{\rm BM} }  {\frac{1}{\delta}}   \\
& \times  \int_{{L_n}}^{{L_{n + 1}}} \int_{{L_n}}^\infty
\frac{\exp \left(  - \frac{{x + y}} {{\Omega _k}( {1 - \rho (\tau _\alpha) } )} \right)} {\Omega _k^2( 1 - \rho (\tau _\alpha)  )} {I_0}\left( \frac{2\sqrt{\rho (\tau _\alpha)}  \sqrt {xy} } { {\Omega _k}( 1 - \rho (\tau _\alpha) )} \right) dy dx
\end{split}
\end{align}
where $ {\cal U}_{s, v} ^{\rm BM}$ denotes the $v$-th user set containing $\delta= |{\cal U}_{s, v} ^{\rm BM}|$ users which all fed back subchannel gain for subchannel $s$ in the previously UL transmission, with $v = 1,2,\ldots, \binom{K}{\delta}$.

\end{theo}


Due to space limitation, we omit the proof of Theorems \ref{theo:3} and \ref{theo:4}, which can be obtained respectively by extending the proof of Theorems \ref{theo:1} and \ref{theo:2} with the same approach used in \cite{Guharoy2013Joint} for analyzing feedback user set. 

\section{Energy Efficiency Analysis}
\label{sect: EE analysis}

In this section, based on throughput analysis, we further study users' EE of the standalone LTE-U network.
%
%
%
It can be calculated by the ratio of the amount of data received from  all subchannels in all subframes to all users' energy consumptions, i.e.,
\begin{align}\label{eq: EE}
\begin{split}
\kappa _{\rm{L}}^{\gamma ,\vartheta }
= \frac{ T_{\rm sb} \frac{B}{S}  \sum\limits_{\alpha  = 1}^{N_{\rm sb}} \sum\limits_{s = 1}^S {\mathbb E}( \eta _{s,\alpha }^{\gamma ,\vartheta } )} { {\mathbb E} (E_0 ) + {\mathbb E} (E_{\rm u} ) + {\mathbb E} (E_{\rm d} )}
= \frac{ T_{\rm sb} B  \sum\limits_{\alpha  = 1}^{{N_{\rm sb}}} {\mathbb E}(\eta _{s,\alpha }^{\gamma ,\vartheta })  }{ {\mathbb E} (E_0 ) + {\mathbb E} (E_{\rm u} ) + {\mathbb E} (E_{\rm d} ) }
\end{split}
\end{align}
where $E_{\rm d}$ and $E_{\rm u}$ represent the energy consumption of all users in a DL data transmission and its previous UL data transmission for reporting CSI, respectively, and $E_0$ denotes the basic circuit energy consumption of all users in the aforesaid whole period. Here, the second equality holds for the same reason for the second equality in (\ref{eq: network throughput}).
%
%
In below, we analyze ${{\mathbb E}\left(E_0\right)}$, ${{\mathbb E}\left(E_{\rm u}\right)}$, and ${{\mathbb E}\left(E_{\rm d}\right)}$ one by one.

For ${{\mathbb E}\left(E_0\right)}$, similar to \cite{yu2015joint}, letting $P_{\rm bs}$ denote per user's basic circuit power consumption in a DL data transmission and its previous UL data transmission, we have
%
\begin{align}\label{eq: Expected energy consumption E ba}
{\mathbb E}\left( E_0 \right) =
K P_{\rm bs} {\mathbb E}\left(t_{\rm t,u} + t_{\rm t,d}\right).
\end{align}

For a UL data transmission, it consists of three periods, the UL channel contention period, the UL reservation period, and the effective UL transmission.
Accordingly, in these periods, all users need to sense the channel, send the reservation signals, and report subchannel gain, according to the adopted  feedback scheme, respectively.
%
The power consumptions for a user in the former two periods are  defined as $P_{\rm se}$ and $P_{\rm rs }$, respectively.
%
Assume that each user transmits with the same fixed power to feed back CSI. Let $\xi (\delta)$ denote the associated energy consumption for feeding back channel quality for $\delta$ subchannels.
According to \cite{3GPP36_213,chiumento2017impact}, $\xi (\delta)$ is an increasing yet non-linear function of $\delta$, due to coding for subchannel index.
Then, expectation ${{\mathbb E}\left(E_{\rm u} \right)}$ satisfies
\begin{align}\label{eq: Expected energy consumption E u}
{\mathbb E}\left( E_{\rm u} \right) =
K (P_{\rm se} {\mathbb E}\left(t_{\rm u}\right) +  {P_{\rm rs}} {\mathbb E} \left(t_{\rm u,r}\right)) + \sum\limits_{k=1}^K  {\mathbb E}\left( E_k^{\vartheta}\right)
\end{align}
where ${\mathbb E}\left(t_{\rm u}\right)$ and ${\mathbb E} \left(t_{\rm u,r}\right)$ are given by
\begin{equation}
{\mathbb E}\left(t_{\rm u}\right) = \sum\limits_{t \ge 0} t p(t)
\end{equation}
%
%
%
\begin{equation}
{\mathbb E} \left(t_{\rm u,r}\right) = {\mathbb E} \left( t_{\rm u,p} \right) -  {\mathbb E} \left( t_{\rm u} \right)
= \sum\limits_{{b_{\rm u}} \in {{\mathbb Z}^ + }} \sum\limits_{t \in \left(( b_{\rm u} - 1) T_{\rm sb}, b_{\rm u} T_{\rm sb} \right]} \left(  b_{\rm u} T_{\rm sb} - t \right) p(t)
\end{equation}
%
%
%
%
and depending on the feedback scheme, energy consumed by user $k$ for CSI feedback, ${\mathbb E}\left( E_k^{\vartheta} \right)$, can be calculated as
\begin{align}\label{eq: Expected E mk}
{\mathbb E}\left( E_k^{\vartheta}\right) =
    \begin{cases}
        \sum\limits_{\delta=1}^{S} \binom{S}{\delta} \exp(-\frac{\lambda \delta }{\Omega_k}) (1- \exp(-\frac{\lambda}{\Omega_k}))^{S-\delta} \xi\left( \delta \right), & \vartheta = {\rm TH} \\
         \xi \left( m \right),              & \vartheta = {\rm BM}.
    \end{cases}
\end{align}

Similarly, for a DL data transmission, it is comprised of the DL channel contention period, the DL reservation period, and the effective DL transmission.
It is noteworthy that to detect and synchronize with the BS each user has to sense the channel in the former two periods with power consumption $P_{\rm se}$.
In the third period, those users scheduled by the BS receive data and the power consumption of receiving on the whole $S$ subchannels is denoted by $P_{\rm rx}$.
In addition, in the last subframe of the effective DL transmission, to obtain the latest CSI each user  estimates channel with power consumption $P_{\rm es}$.
Then, the expected energy consumption ${{\mathbb E}\left(E_{\rm d}\right)}$ can be decomposed as
\begin{align}\label{eq: Expected energy consumption E d}
{\mathbb E}\left( E_{\rm d} \right) =
K P_{\rm se} {\mathbb E}\left(t_{\rm d,p}\right) + {\mathbb E} \left(E_{\rm rx}^\vartheta\right)  +  K{P_{\rm es}} {T_{\rm sb}}
\end{align}
where ${\mathbb E} \left(t_{\rm d,p}\right) = \sum\nolimits_{b_{\rm d} \in {{\mathbb Z}^ + }}  b_{\rm d} T_{\rm sb} 
\sum\nolimits_{t \in \left((b_{\rm d} - 1) T_{\rm sb},  b_{\rm d} T_{\rm sb}\right]} p(t) $
measures the average time of the DL pre-transmission period,
and $E_{\rm rx}^\vartheta$ denotes the energy consumption for those users receiving data in the effective DL transmission.

For each subchannel in a subframe, only the user allocated this subchannel receives data.
The user's energy consumption in this duration is ${{P_{\rm rx}} {T_{\rm sb}}} / S$ \cite{yu2015joint}.
Taking into account $N_{\rm sb}$ subframes each with $S$ subchannels, the energy consumption of all users in an effective DL transmission is
\begin{align}\label{eq: ECD Transmission}
{\mathbb E}  \left( E_{\rm rx}^\vartheta\right)  = \frac{ P_{\rm rx} T_{\rm sb} }{S}  \sum\limits_{\alpha  = 1}^{ N_{\rm sb} }  \sum\limits_{s = 1}^S  {\mathbb E} \left( I\left(  N_s^\vartheta  \ge 1 \right)\right)  = P_{\rm rx}T_{\rm sb}N_{\rm sb} {\mathbb E}\left(I\left(N_s^\vartheta \ge 1\right)\right)
\end{align}
where $N_s^\vartheta$ is the number of users fed back CSI for subchannel  $s$ under feedback scheme $\vartheta$. $I\left(\cdot\right)$ is an indictor function, which equals 1 if $N_s^\vartheta \ge 1$, otherwise equals 0.
%
${\mathbb E}\left(I\left(N_s^\vartheta \ge 1\right)\right)$ can be further decomposed as follows:
\begin{itemize}
\item For $\vartheta = { \rm TH}$, when $N_s^\vartheta \ge 1$, there exists at least one user of which the channel gain of subchannel $s$ is not less than $\lambda$. Therefore,
%
${\mathbb E}\left({I\left( {N_s^{{\rm TH}} \ge 1} \right)}\right)= 1- \prod\limits_{k=1}^K  \int\nolimits_0^\lambda  \frac{1}{\Omega_k} \exp \frac{g}{\Omega_k}dg  =1 - \prod\limits_{k = 1}^K {\left[ {1 - {\rm exp}\left( { - \frac{\lambda }{{{\Omega _k}}}} \right)} \right]}.$

\item For $\vartheta = { \rm BM}$, when $N_s^\vartheta \ge 1$, there exists at least one user of which subchannel $s$ is one of its best $m$ subchnannels. Therefore,
%
${\mathbb E}\left({{I\left( {N_s^{\rm BM} \ge 1} \right)}} \right)= 1 - {\left( {\frac{{S - m}}{S}} \right)^K}.$

\end{itemize}

\section{Energy Efficiency Optimization}
\label{sect: EE optimization}

In this section, based on the analysis in Section \ref{sect: EE analysis} we maximize users' EE of a standalone LTE-U network under different feedback schemes.
For the frequency-domain scheduler, we use the greedy scheduler as an example due to its benefits of exploiting multi-user diversity thus to improve  network throughput.
Specifically, to guarantee Wi-Fi's performance, we firstly analyze the time ratio that Wi-Fi can successfully occupy channel when coexisting with the standalone LTE-U network. Then, coexistence-aware users' EE optimization for the standalone LTE-U network is studied.

\subsection{Coexistence Awareness}
\label{subsect: Fairness Coexistence}


Let $t_{\rm s}^{\rm w}$ denote Wi-Fi's channel occupancy time ratio when coexisting with LTE-U.
Given both Wi-Fi and LTE-U networks' loads and MAC protocol settings, similar to \cite{Bianchi2000Performance} $t_{\rm s}^{\rm w}$, defined mathematically as the fraction of successful packet transmission time, can be written as
\begin{align}\label{eq: time ratio of Wi-Fi}
t_{\rm s}^{\rm w} =\frac{p_{\rm T} p_{\rm s}^{\rm w} T_{\rm s}^{\rm w}} {\left( 1 - {p_{\rm T}} \right)\sigma  + p_{\rm T} p_{\rm s}^{\rm w} T_{\rm s}^{\rm w} + p_{\rm T} \left( 1 - p_{\rm s }^{\rm w} \right)T_{\rm c}^{\rm w}}
\end{align}
where $T_{\rm s}^{\rm w}$  is the duration of a successful  Wi-Fi transmission, $p_{\rm T}$ and $p^{\rm w}_{\rm s}$ given as follows are the probability that at least a Wi-Fi STA or the standalone LTE-U network occupies the channel and the probability that a Wi-Fi STA successful sends a packet, respectively,
\begin{align}\label{eq: Pro one node tx}
{p_{\rm T}} = 1 - {\left( {1 - {\tau _{\rm w}}} \right)^{{N_{\rm w}}}}\left( {1 - {\tau _{\rm L}}} \right)
\end{align}
\begin{align}\label{eq: Pro wifi success}
p_{\rm s}^{\rm w} = {N_{\rm w}}{{\left( {1 - {\tau _{\rm w}}} \right)}^{{N_{\rm w}} - 1}}\left( {1 - {\tau _{\rm L}}} \right)/p_{\rm T}
\end{align}
with $\tau_{\rm w}$ and $\tau_{\rm L}$  representing the probability of a Wi-Fi STA sending a packet and that of the BS or the users sending reservation signals in any backoff slot, respectively.
Here, $\tau_{\rm w}$ and $\tau_{\rm L}$ can be figured out numerically by jointly solving (\ref{eq: tau w})-(\ref{eq:  pL}) in Appendix \ref{app:deduce tau pL}.

To share channel fairly, let $D_{\rm th}$ denote the minimal channel occupancy time ratio that Wi-Fi needs. If $t_{\rm s}^{\rm w} \ge D_{\rm th}$, we consider that Wi-Fi's performance is guaranteed. 
For the value of $D_{\rm th}$, we assume that it can be negotiated by the two networks at some signaling costs.

\subsection{EE under Threshold-based Feedback Scheme}
\label{subsect: EE thre}
To achieve coexistence-aware users' EE maximization for the standalone LTE-U network, we study the following optimization problem
\begin{subequations}
\label{eq:orginal problem 1}
\vspace{-0.2cm}
\begin{eqnarray}
\label{original 1 objective function}
\text{OP1:}~~
\max\limits_{\lambda, Z}  \quad &&
\kappa _{\rm L}^{\rm G , TH }\left(\lambda,Z\right) = \frac{ T_{\rm sb} B \sum\limits_{\alpha  = 1}^{N_{\rm sb} } {\mathbb E} \left(\eta _{s,\alpha }^{\rm G ,TH }\right) } { {\mathbb E} (E_0 ) + {\mathbb E} (E_{\rm u} ) + {\mathbb E} (E_{\rm d})  } \\
{\text{s.t.}}\quad
&&\label{eq: org1 fairness const} t^{\rm w}_{\rm s} \ge D_{\rm th}\\
&&\label{eq: org1 Z}  \lambda \ge 0, Z \in {\mathbb Z}^{+}
\end{eqnarray}
\end{subequations}
where (\ref{original 1 objective function}) is the energy efficiency of all users in the standalone LTE-U network adopting the greedy scheduler and threshold-based feedback, (\ref{eq: org1 fairness const}) guarantees Wi-Fi sufficient channel occupancy time, (\ref{eq: org1 Z}) specifies the ranges of feedback threshold and CW size of the standalone LTE-U network.
It is noteworthy that, $t^{\rm w}_{\rm s}$ varies with $Z$, as both networks coexist in the same channel.

As the relaxed version of the objective function is non-convex and non-concave and variable $Z$ is  integer, directly solving (\ref{eq:orginal problem 1}) is difficult. In below, we adopt a decomposition based approach, in which the original OP is decomposed into two subproblems each with a single variable:
\begin{subequations}
\label{eq:op1 sb1}
\begin{eqnarray}
\label{org 1 sub1 OB}
\text{SOP1:}~~ \max\limits_{ Z}  \quad &&
 \kappa _{\rm L}^{\rm G , TH }\left(Z\right) 
\\
{\text{s.t.}}\quad
&&\label{eq: org1 sb1 fairness const} t^{\rm w}_{\rm s} \ge D_{\rm th}\\
&&\label{eq: org1 sb1 Z} Z \in {\mathbb Z}^{+}
\end{eqnarray}
\vspace{-1cm}
\end{subequations}
\begin{subequations}
\label{eq:op1 sb2}
\begin{eqnarray}
\label{org 1 sub2 OB}
\text{SOP2:}~~\max\limits_{\lambda }  \quad &&
 \kappa _{\rm L}^{\rm G , TH }\left(\lambda \right) 
 \\
{\text{s.t.}}\quad
&&\label{eq: org1 sb2 threshold const} \lambda \ge 0.
\end{eqnarray}
\end{subequations}

\subsubsection{Solution to SOP1}
\label{subsect: SOP1}
In general, it is difficult to solve SOP1 optimally with theoretical approaches, because to evaluate $ \kappa _{\rm L}^{\rm G , TH }\left(Z\right)$ one needs to calculate probabilities $P\{ \tau _\alpha = a T_{\rm sb} \}$ and ${\bar p_{{\rm c},\alpha }}$ (also functions of $Z$) numerically based on Lattice-Poisson algorithm (see Section \ref{sect: throughput analysis}).
Though exhaustive search can find the optimal solution to SOP1, it incurs high computational complexity also due to frequently invoking Lattice-Poisson algorithm. 
However, in below we argue that the minimum CW size satisfying (\ref{eq: org1 sb1 fairness const}) and (\ref{eq: org1 sb1 Z}) offers a suboptimal yet fine solution to the problem.

\begin{algorithm}
\caption{Threshold Searching Algorithm}
\label{alg: algorithm1}
\begin{algorithmic}[1]
\algtext*{EndFunction}
\State {Set tolerance $\beta >0$, maximum threshold $\lambda_{\rm max}>0$, $\lambda^* = 0$;}
\For {$j =1$ to $J$}
    \State  {Initialize a starting point $\lambda_j \in \left[0, \lambda_{\rm max} \right]$;}
    \State {\textbf{repeat}}  \Comment {Gradient descent method}

            \State {\quad  $\Delta \lambda_j \leftarrow  d \kappa _{\rm L}^{\rm G, TH}( \lambda _j )/{d\lambda_j}$};
                  \Comment {Compute the negative gradient}

        \State {\quad Break if $\left|   \Delta \lambda_j  \right| \le \beta $;} \Comment {Stopping criterion}

        \State {\quad$ l \leftarrow {\rm argmin}{_{\upsilon  \ge 0}} -\kappa_{\rm L}^{\rm G,TH}( {\lambda_j  + \upsilon \Delta \lambda_j } )$;} \Comment {Choose step size by line search}

        \State {\quad Update $\lambda_j \leftarrow \lambda_j + l \Delta \lambda_j$;}

  \State{Obtain $\lambda^*_j = \lambda_j$;}

  \State { ${\lambda ^*} =  \arg \min_{\lambda^* _j} \left\{ -\kappa_{\rm L}^{\rm G,TH} \left( \lambda^* _j  \right),-\kappa_{\rm L}^{\rm G,TH} \left( \lambda^* \right) \right\} $};
  \Comment {Compare to find an optimized result}

\EndFor


\end{algorithmic}
\end{algorithm}

Notice that with a greater CW size $Z$ comes a higher Wi-Fi's channel occupancy time ratio.
So, constraint (\ref{eq: org1 sb1 fairness const}) can be transformed into $Z \ge Z_{0}$, where $Z_0$ is the minimum CW size guaranteeing Wi-Fi's channel occupancy time ratio no less than $D_{\rm th}$ and can be calculated numerically from (\ref{eq: time ratio of Wi-Fi})-(\ref{eq: Pro wifi success}) and (\ref{eq: tau w})-(\ref{eq: pL}).
Then, solving SOP1 is equivalent to addressing $\arg \max \limits_{Z \ge Z_{0}, Z \in {\mathbb Z}^{+}} \kappa _{\rm L}^{\rm G , TH }\left(Z\right)$.

For the denominator of $\kappa _{\rm L}^{\rm G , TH }\left(Z\right)$  (see (\ref{original 1 objective function})),
only ${\mathbb E} \left(t_{\rm u}\right)$ and ${\mathbb E} \left(t_{\rm u,r}\right)$ in ${\mathbb E} \left(E_{\rm u}\right)$ (see (\ref{eq: Expected energy consumption E u})),
${\mathbb E} \left(t_{\rm d,p}\right)$ in ${\mathbb E} \left(E_{\rm d}\right)$ (see (\ref{eq: Expected energy consumption E d})),
and ${\mathbb E} \left( t_{\rm t,u} + t_{\rm t,d} \right)$ in ${\mathbb E} \left(E_0\right)$  (see (\ref{eq: Expected energy consumption E ba}))
depend on $Z$.
As UL reservation period $t_{\rm u,r}$ is much shorter than $t_{\rm t,u}$, we neglect the energy consumed in $t_{\rm u,r}$.
Also, notice that, 
due to more backoff counter choices available,
${\mathbb E}\left( t_{\rm u} \right)$, ${\mathbb E}\left( t_{\rm d,p} \right)$, and ${\mathbb E}\left( t_{\rm t,d} + t_{\rm t,u} \right)$ increase with  $Z$.
So, the denominator of $\kappa _{\rm L}^{\rm G , TH }\left(Z\right)$ can be approximated as an increasing function of $Z$.
%
%
On the other hand, for the numerator of $\kappa _{\rm L}^{\rm G , TH }\left(Z\right)$, strictly speaking, though $ {\mathbb E}(\eta _{s,\alpha }^{\rm G,TH})$ can be a non-decreasing function of $Z$,
${\mathbb E} (\eta _{s,\alpha }^{{\rm G, TH}}| \tau _\alpha)$ is a decreasing function of feedback delay $\tau_\alpha$ \cite{Guharoy2013Joint} and
%
the mean of $\tau_\alpha$ increases with $Z$.
So, taking into account the characteristic of the denominator of $\kappa _{\rm L}^{\rm G , TH }\left(Z\right)$, it could be a suboptimal yet fine solution to set $Z^* = Z_0$, which is a choice independent of $\lambda$ but allows LTE-U to behave as aggressively as possible yet still leave enough channel occupancy time for Wi-Fi.
The performance of the solution  will be studied in Section \ref{subsect: Parameter Settings}.



\subsubsection{Solution to SOP2}
\label{subsect: SOP2}
For SOP2, we have the following  proposition.

%

\begin{prop} \label{Energy with lambda}
Both the numerator and denominator of $\kappa _{\rm L}^{\rm G , TH }\left(\lambda \right)$ are non-convex and non-concave functions.
\end{prop}

We omit the proof due to space limitation. But the proposition can be easily checked by setting $N_{\rm w} = 6$, $Z = 64$, $K=5$ and the other parameters as used in Section \ref{subsect: Parameter Settings}.

%

It is difficult to directly solve the optimization problem \cite{Crouzeix1991Algorithms, Barros1996A}.
So, we propose a suboptimal algorithm as illustrated in Algorithm 1 to address SOP2.
The algorithm is designed simply by invoking gradient descent method \cite{Boyd2014convex} for $J$ times (steps 3 - 8), each with a new initiator $\lambda_j^{(0)}$.
Each suboptimal result found by a new search is recorded in $\lambda_j^*$ (step 9).
The best search result is obtained by comparing the energy efficiency of all found $\lambda_j^*$'s (step 10).

%

In short, to address (\ref{eq:orginal problem 1}), one can first find a preferred CW size $Z^*$ from SOP1.
Then, substituting $Z$ in (\ref{original 1 objective function}) with $Z^*$ and using Algorithm \ref{alg: algorithm1}, an optimized $\lambda^*$ is derived.

%
%

\subsection{EE under Best-$m$ Feedback Scheme}

Similarly, when the standalone LTE-U network adopts the greedy scheduler and best-$m$ feedback, the problem of coexistence-aware users' EE maximization can be formulated as
\begin{subequations}
\label{eq:orginal problem 2}
\begin{eqnarray}
\label{original 2 objective function}
\text{OP2:}~~
\max\limits_{m, Z}  \quad &&
\kappa _{\rm L}^{\rm G , BM }\left(m,Z\right) \\
{\text{s.t.}}\quad
&&\label{eq: org2 fairness const} t^{\rm w}_{\rm s} \ge D_{\rm th} \\
&&\label{eq: org2 BM m} 0 \le m \le S\\
&&\label{eq: org2 BM mZ}m \in {\mathbb Z}, Z \in {\mathbb Z}^{+}.
\end{eqnarray}
\end{subequations}

As (\ref{original 2 objective function}) is an integer OP and the relaxed objective function is non-convex and non-concave,  to solve OP2 we can still decompose it into two subproblems each with a single variable. 
%
Similar to SOP1, for $Z_0$ the minimum CW size guaranteeing Wi-Fi's channel occupancy time ratio no less than $D_{\rm th}$ is still a suboptimal yet fine solution. For the optimal $m$, as it is an integer and $S$ is usually far from large in a practical system, the optimal value $m^*$ can be found simply via exhaustive search.


\section{Simulation Results}
\label{sect: simulation results}

In this section, we first verify the theoretical analysis of the network throughput and users' EE of the standalone LTE-U network under different scheduler-feedback scheme combinations. Then we evaluate the proposed coexistence-aware EE optimization algorithms to get insight into the relationship between users' EE performance and fairness guarantee for Wi-Fi.

\subsection{Parameter Settings}
\label{subsect: Parameter Settings}

We consider that a standalone LTE-U network serving 10  users coexists with a Wi-Fi network in a 20 MHz channel with center frequency at 5.75 GHz.
Each user moves in the LTE-U network with a fixed speed of 3 km/h.
In the standalone LTE-U network, the channel is divided into 20 subchannels.
The subframe length in the LTE-U network $T_{\rm sb}$ and the number of successive subframes $N_{\rm sb}$ it occupies channel for DL/UL transmission are set as 1 ms and 3, respectively.
The CW size of the LTE-U network $Z$ is set as 64.
%
The transmit power of a single user, the power consumption for channel estimation $P_{\rm es}$, the power consumption for sensing channel $P_{\rm se}$, the power consumption for sending reservation signal $P_{\rm rs}$, the power consumption for receiving data $P_{\rm re}$,
and the power consumption for basic circuit $P_{\rm bs}$ are 1 W, 200 mW, 11 mW, 100 mW, 200 mW, and 0.1 mW, respectively.
Here, as all 10 users can transmit reservation signal on orthogonal subchannels simultaneously, we assume  each user transmits reservation signal in a low-transmit power mode of 100 mW.
According to \cite{chiumento2017impact}, energy consumption model for CSI feedback is set as $\xi\left( \delta \right) =(4+2 \delta + \left\lceil \log_2 \binom{100}{\delta} \right\rceil )e_0$, where $e_0$ is the energy consumed for sending one bit information.
Given the transmit power of a single user of 1 W and assuming that for reliable feedback CSI information is sent with the lowest order modulation QPSK and coding efficiency 0.1523 of the LTE system\cite{chiumento2017impact}, $e_0$ can be found out as $2.28 \times 10^{-6}$ J.
The data rate in the standalone LTE-U network supports the fifteen different rates $r_{n}$ defined in \cite{3GPP36_213}, $n= 1, 2, ..., 15$.
The associated thresholds for discrete rate adaptation are set according to the approach in \cite{Baum2003Performance}: ${r_n} = {\rm log}{_2}\left( {1 + \varsigma {L_n}} \right)$, where $\varsigma = 0.398$ denotes the coding loss of a practical code.
For users' subchannel gain, in the i.i.d. user scenario, all users have the same mean subchannel gain set as $\Omega = 7.78$ dB; in the non-i.i.d. user scenario, the mean subchannel gain of user $k  \in {\cal K} $ is set with the same approach in \cite{Guharoy2013Joint} as $\Omega_k = \Omega \mu^{k-1}$, where $\mu >0$. In the simulation, we set $\mu = 1.1$.
The parameters of the Wi-Fi network follows the IEEE 802.11ac standard \cite{Ong2011IEEE}.
The duration of a backoff slot $\sigma$, the maximum backoff stage of Wi-Fi $b_{\rm w}$, and the minimum CW of Wi-Fi $W$ are 9 us, 5, and 32, respectively.
Similar to \cite{Li2017Modelling}, in the simulation we set the duration of a successful Wi-Fi transmission $T^{\rm w}_{\rm s}$ and the duration of a collided Wi-Fi transmission $T^{\rm w}_{\rm c}$ as 540.72 us and 284.72 us, respectively.
For all simulations, we present simulation results by averaging over 5000 DL and UL effective transmission pairs. 

\subsection{Throughput in the i.i.d. User Scenario}
\label{subsect: throughput iid users}

\begin{figure*}
  \centering
  \subfigure[]
             {\label{fig:througput TH}%
             \includegraphics[width=0.49\textwidth]{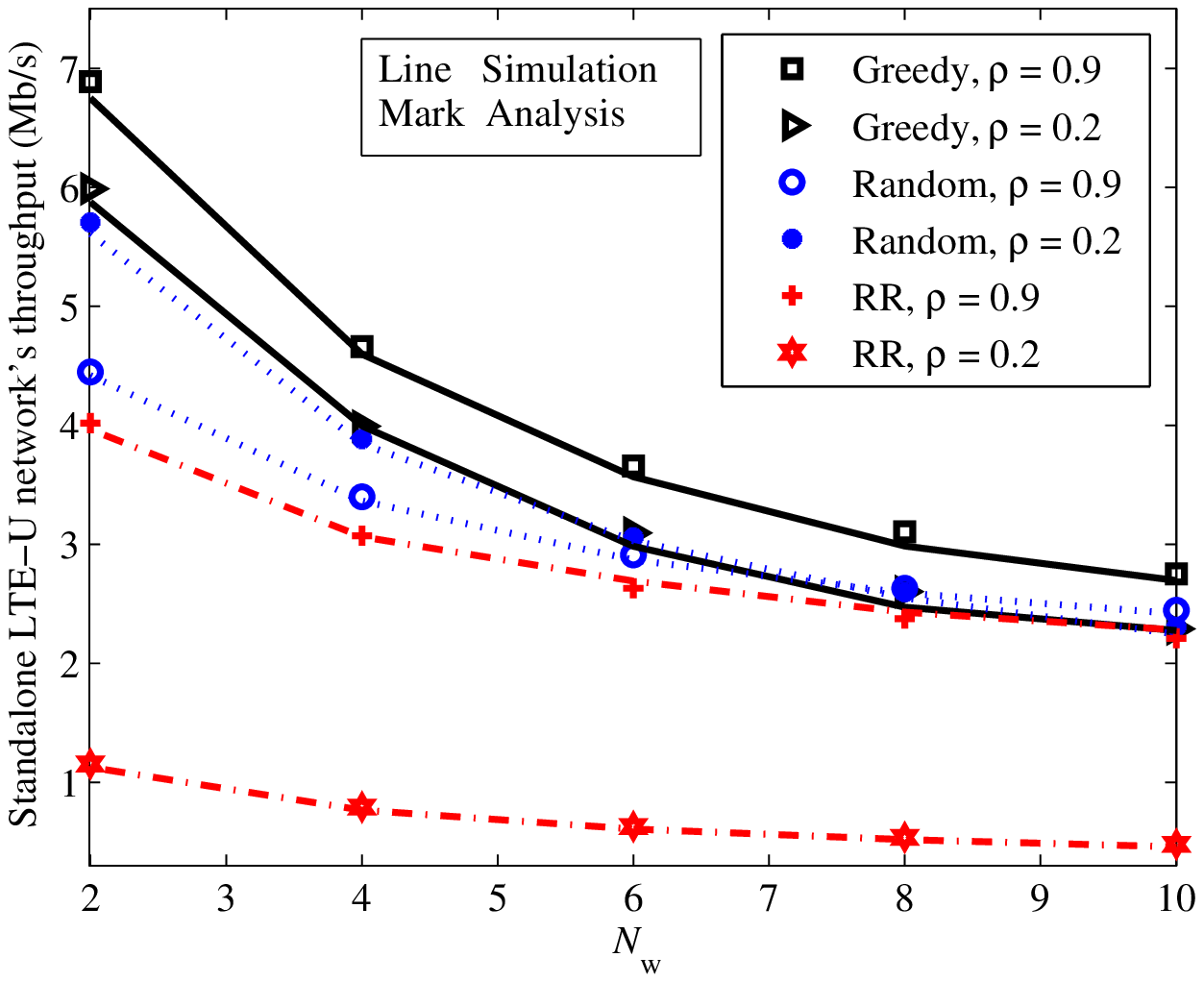}}
  \subfigure[]
             {\label{fig:decrement TH}%
             \includegraphics[width=0.49\textwidth]{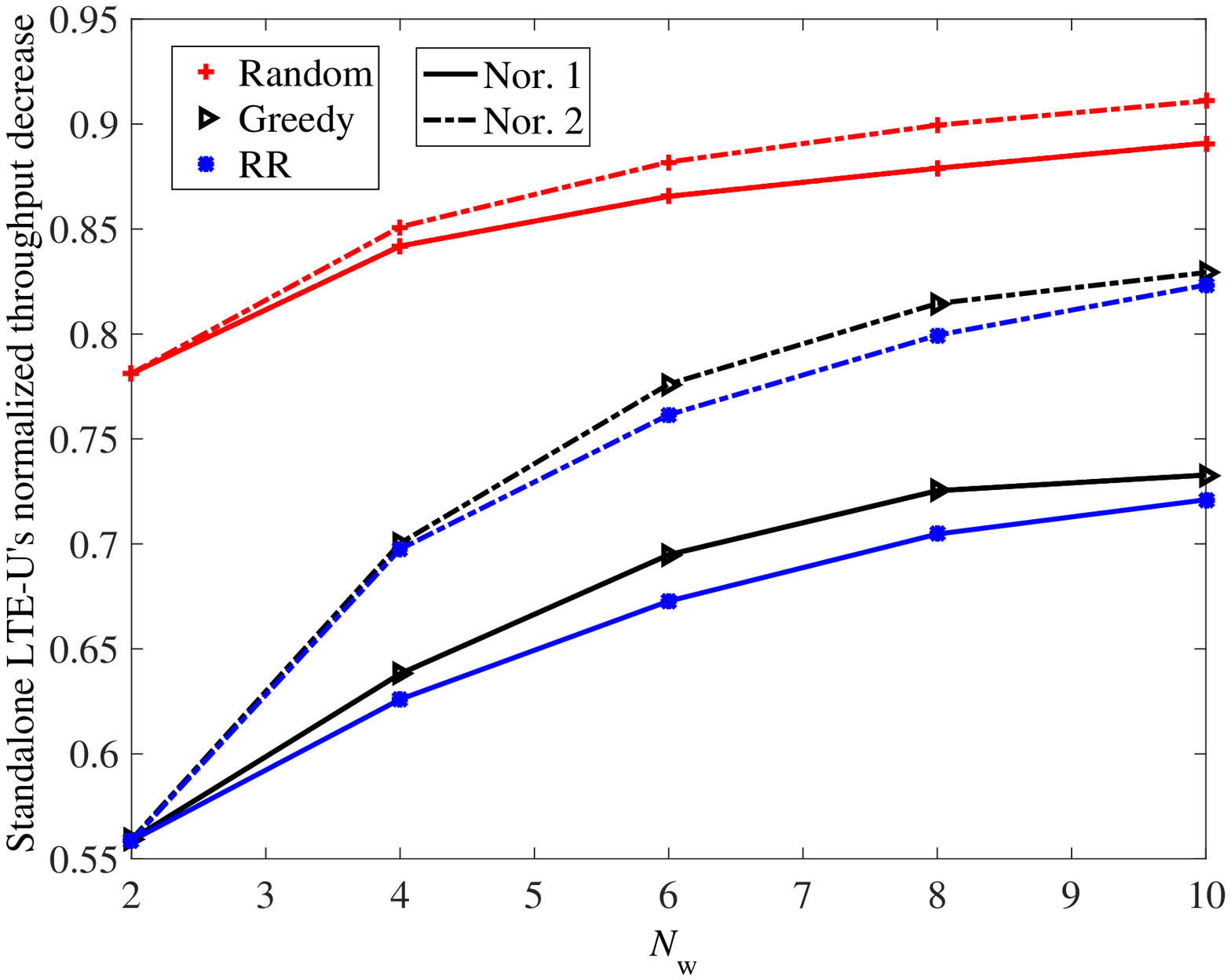}}
             \vspace{-0.15cm}
  \caption{Standalone LTE-U network's throughput for i.i.d. users under the threshold-based feedback scheme. (a) Standalone LTE-U network's throughput. (b) Standalone LTE-U network's normalized throughput decrease at $\rho = 0.2$.}
  \label{fig:throughput and decrement TH}
  \vspace{-0.75cm}
\end{figure*}

In Fig. \ref{fig:throughput and decrement TH}, we study the throughput of the standalone LTE-U network in the i.i.d. user scenario under the threshold-based feedback scheme. 
Notice that for the i.i.d. user scenario, the greedy scheduler and the PF scheduler are equivalent, as the user with the largest subchannel gain is the one with largest normalized subchannel gain. So, here we mainly compare the performance of the round robin, greedy, and random schedulers.
Besides, to show the impact of threshold, we define $\rho = {\exp} \left( { - {\lambda  / \Omega }} \right)$, which gives the probability of a user feeding back CSI for a subchannel, given threshold and mean subchannel gain equal to $\lambda$ and $\Omega$, respectively.
%
%
Obviously, the larger the value of $\lambda$, the smaller the value of $\rho$ and thus the smaller number of users that will feed back CSI for a subchannel.

It is observed from Fig. \ref{fig:througput TH} that the analytical results match closely with the simulation results.
Furthermore, for the greedy scheduler, the standalone LTE-U network's throughput at $\rho = 0.9$ is larger than that at $\rho = 0.2$.
This is because the larger $\rho$ makes more users feed back their subchannel gain, generating chances to utilize channel when the channel quality in general is poor.
However, for the random scheduler, in the test setting the smaller $\rho$ contributes to the better throughput when the Wi-Fi STA number $N_{\rm w}$ is smaller than 8.
This is rational because when $N_{\rm w}$ is small, channel access delay for the LTE-U network is usually  small accordingly, leading to more opportunities to keep the reported CSI not out-of-date.
In such a case, it is beneficial for network throughput if enhancing $\lambda$ thus allowing only users of good channel quality to be randomly scheduled.
%
%
In contrast, for the RR scheduler, the larger $\rho$ always contributes to the higher network throughput.
The reason is similar to that for the greedy scheduler. However, in this case the larger $\rho$ makes only the one user served in order have more chances to feed back its subchannel gain, generating more opportunities to utilize channel for throughput improvement.
The aforesaid finding on the greedy and RR schedulers are consistent with the results in  \cite{Guharoy2013Joint}.
It is also noticed from Fig. \ref{fig:througput TH} that with an increase of Wi-Fi STA number $N_{\rm w}$, the network throughput under each of the three schedulers decreases.
To get more insights into the throughput degradation, in Fig. \ref{fig:decrement TH} we compare network throughput results with two normalization methods (denoted as ``nor. 1" and ``nor. 2") at $\rho = 0.2$.
%
%
The solid lines (normalization method 1) measure the normalized decrease of network throughput $\eta _{\rm L}^{\gamma, \vartheta}(N_{\rm w})$ (as a function of the Wi-Fi STA number $N_{\rm w}$) compared with $\eta _{\rm pri}^{\gamma, \vartheta}(N_{\rm w})$, which is the network throughput when the BS has all users' instant subchannel gains perfectly.
The dotted lines (normalization method 2) show the normalized decrease of throughput $\eta _{\rm L}^{\gamma, \vartheta}(N_{\rm w})$ compared with  $\eta _{\rm pri}^{\gamma, \vartheta }(2)$, which is the network throughput when $N_{\rm w} = 2$ and the BS has all users' instant subchannel gains perfectly.
%
%
%
From Fig. \ref{fig:decrement TH} we can see that with either normalization methods throughput decrease becomes larger as the Wi-Fi STA number (i.e., the Wi-Fi network's traffic load) increases.
For normalization method 1, as  throughput is compared under the same Wi-Fi STA number, the main reason for the increase of throughput decrease with the Wi-Fi STA number is the more and more inaccurate CSI used for rate adaptation. Notice that the channel access delay of the LTE-U network and thus the feedback delay increase as the number of contending Wi-Fi STAs increases.
However, for normalization method 2, as $\eta _{\rm L}^{\gamma, \vartheta}(N_{\rm w})$ is compared with  $\eta _{\rm pri}^{\gamma, \vartheta }(2)$ of fixed Wi-Fi STA number 2, the reasons for the increase of throughput decrease are not only imperfect CSI but also reduced channel occupancy time of the LTE-U network with the Wi-Fi STA number.
So, comparing the gap between the two normalization methods and the throughput decrease calculated with normalization method 1, one shall notice that the imperfect CSI due to feedback delay plays a key role in deteriorating the performance of channel-aware transmission of the standalone LTE-U network.


\begin{figure*}
  \centering
  \subfigure[]
             {\label{fig:througput BM} %
             \includegraphics[width=0.49\textwidth]{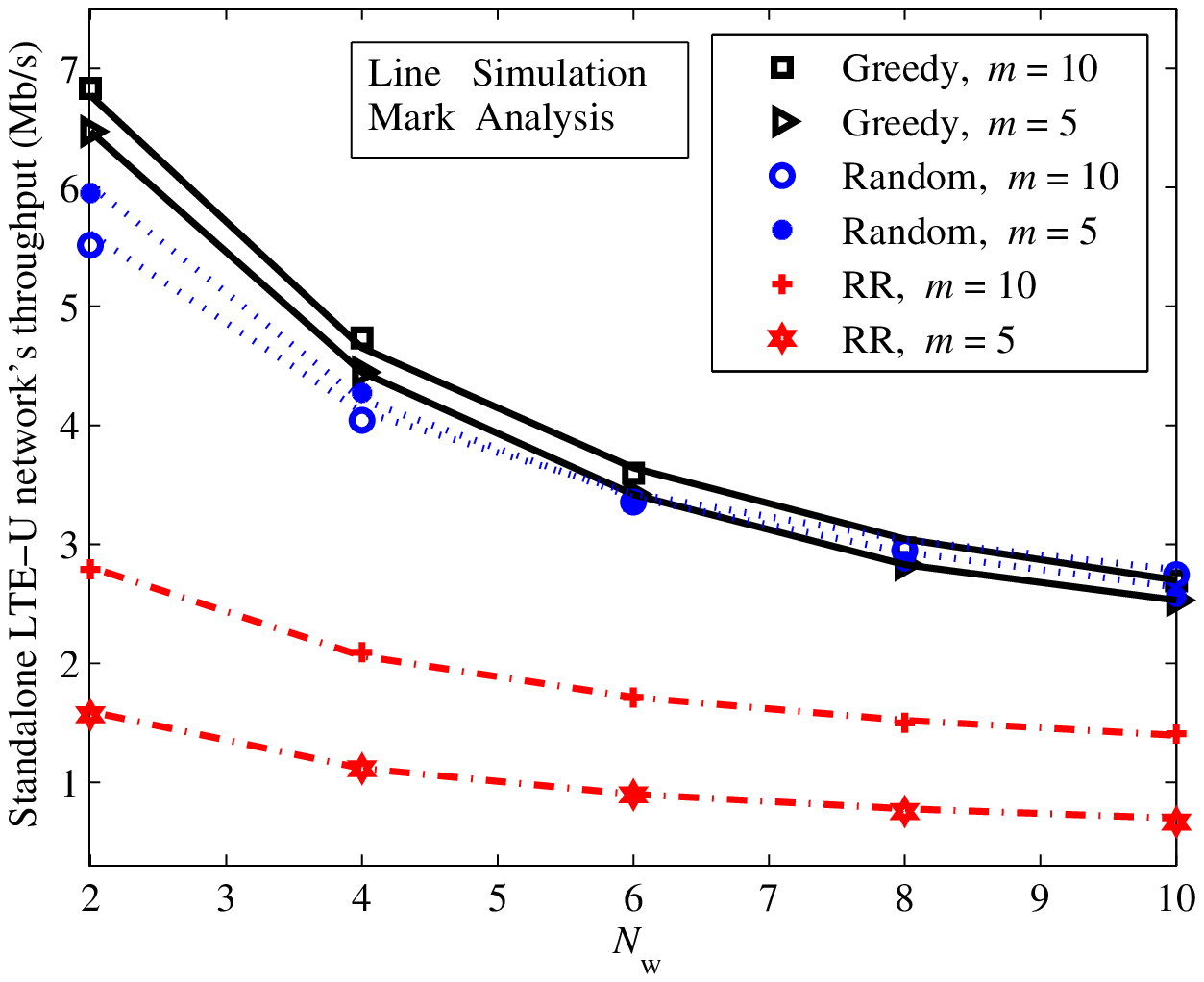}}
  \subfigure[]
             {\label{fig:decrement BM}%
             \includegraphics[width=0.49\textwidth]{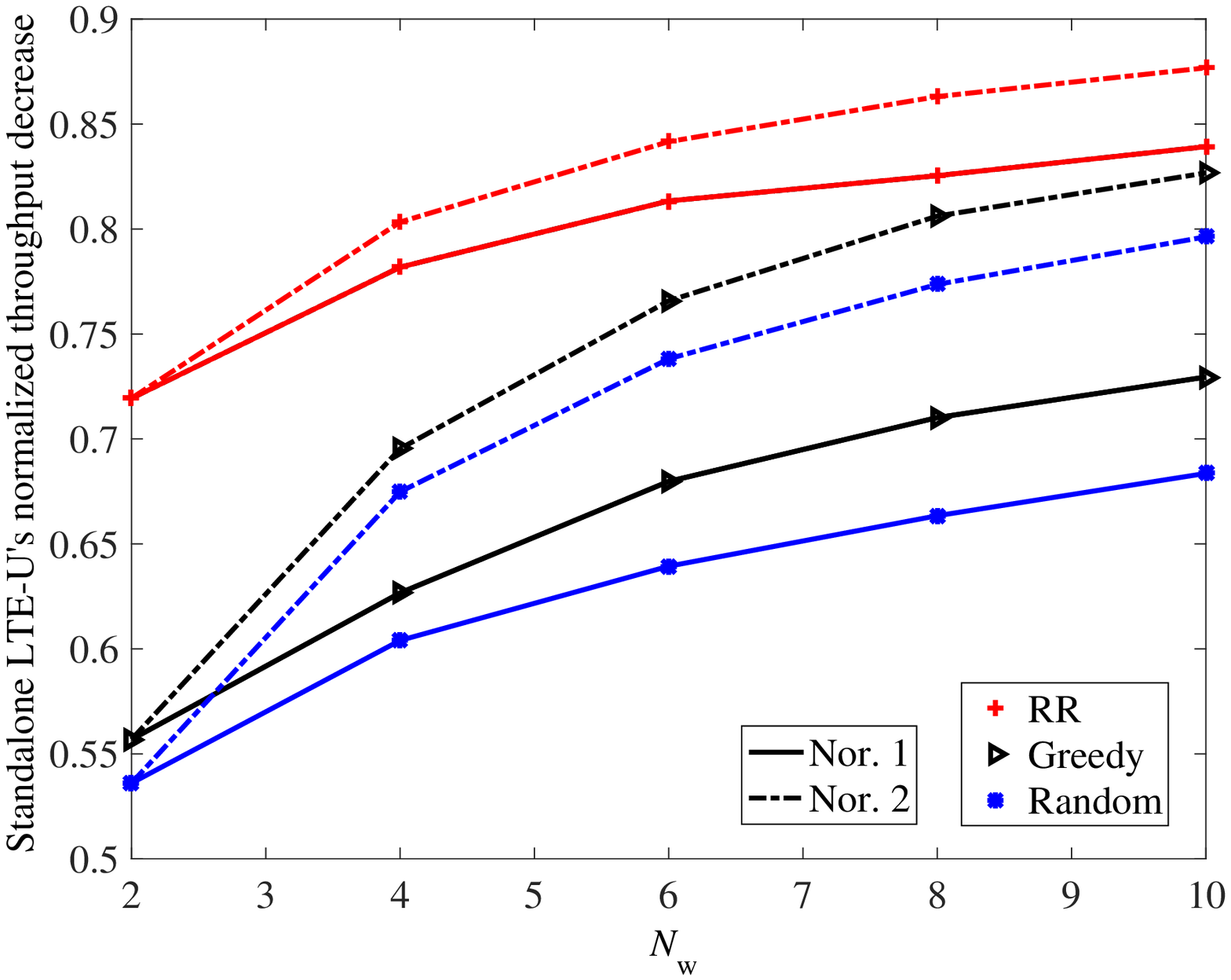}}
             \vspace{-0.35cm}
  \caption{Standalone LTE-U network's throughput for i.i.d. users under the best-$m$ feedback scheme. (a) Standalone LTE-U network's throughput. (b) Standalone LTE-U network's normalized throughput decrease at $m = 5$.}
  \label{fig:throughput and decrement BM}
  \vspace{-0.75cm}
\end{figure*}

Fig. \ref{fig:throughput and decrement BM} shows the network throughput of the standalone LTE-U network in the i.i.d. user scenario  under the best-$m$ feedback scheme.
As increasing $m$ has the similar effect of increasing $\rho$, for each scheduler, as compared with results between $\rho$ equal to 0.9 and 0.2 similar results between $m$ equal to 10 and 5 are observed.
Yet, here one notices clearer that the greedy scheduler does not always work best.
For instance, when $N_{\rm_W}= 10$, the random scheduler works better than the greedy scheduler. This is because the former choose user to serve randomly from multiple candidates all reported good channel quality, which is a more robust strategy to deal with long feedback delay issue.

%
%

\subsection{Throughput in the Non-i.i.d. User Scenario}
\label{subsect: throughput noniid users}

\begin{figure*}
  \centering
  \subfigure[]
            {\label{fig:througput TH NIID}%
             \includegraphics[width=0.49\textwidth]{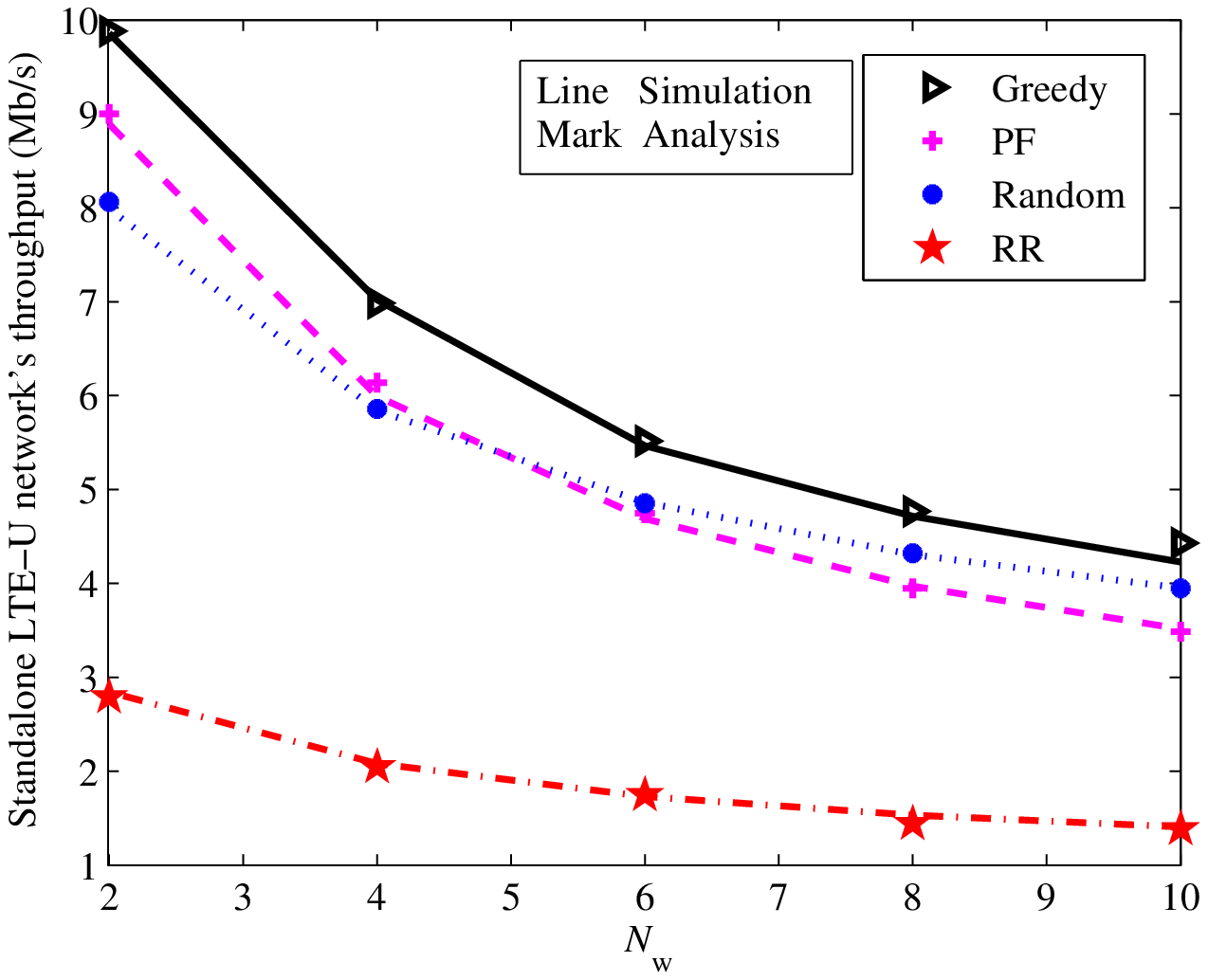}}
  \subfigure[]
            {\label{fig:decrement BM NIID}%
            \includegraphics[width=0.49\textwidth]{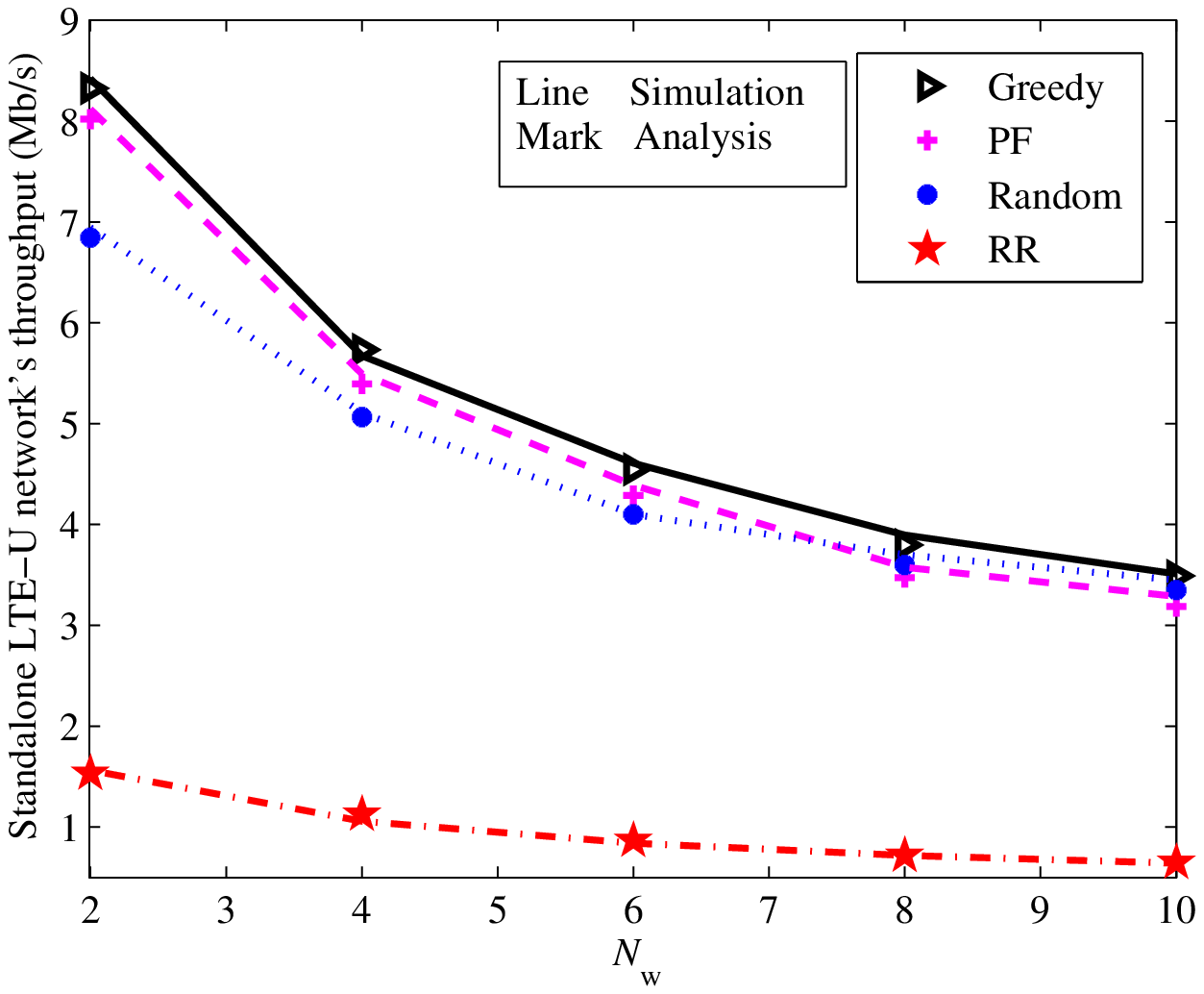}}
            \vspace{-0.15cm}
  \caption{Standalone LTE-U network's throughput for non-i.i.d. user scenario. (a) Throughput under the threshold-based feedback scheme at $\rho = 0.5$. (b) Throughput under the best-$m$ feedback scheme  at $m = 5$.}
  \label{fig:throughput NIID}
  \vspace{-0.75cm}
\end{figure*}

Fig. \ref{fig:throughput NIID} shows the performance of the four schedulers in the non-i.i.d. user scenario under both the threshold-based feedback scheme and best-$m$ feedback scheme. Here, for the threshold-based feedback scheme we set $\rho = 0.5$, while for the best-$m$ feedback scheme we set $m = 5$.
Again, we  find that the analytical results match well with the simulation results.
Further, it is observed that without using CSI information to schedule users, the performance of the round robin scheduler is much worse than other three schedulers.
As expected the greedy scheduler is always better than the PF scheduler; however, the performance gap between the random scheduler and the greedy scheduler reduces with the number of Wi-Fi STAs, for the same reason as we discussed for Fig. \ref{fig:throughput and decrement BM}.
Moreover, it is found that the random scheduler offers higher throughput than the PF scheduler when the Wi-Fi STA number is larger than 5 (resp. 7) for the threshold-based (resp. best-$m$) feedback scheme.


\subsection{Users' EE of the Standalone LTE-U Network}

\begin{figure*}
  \centering
  \subfigure[]%
            {\label{fig:EE TH IID}%
             \includegraphics[width=0.49\textwidth]{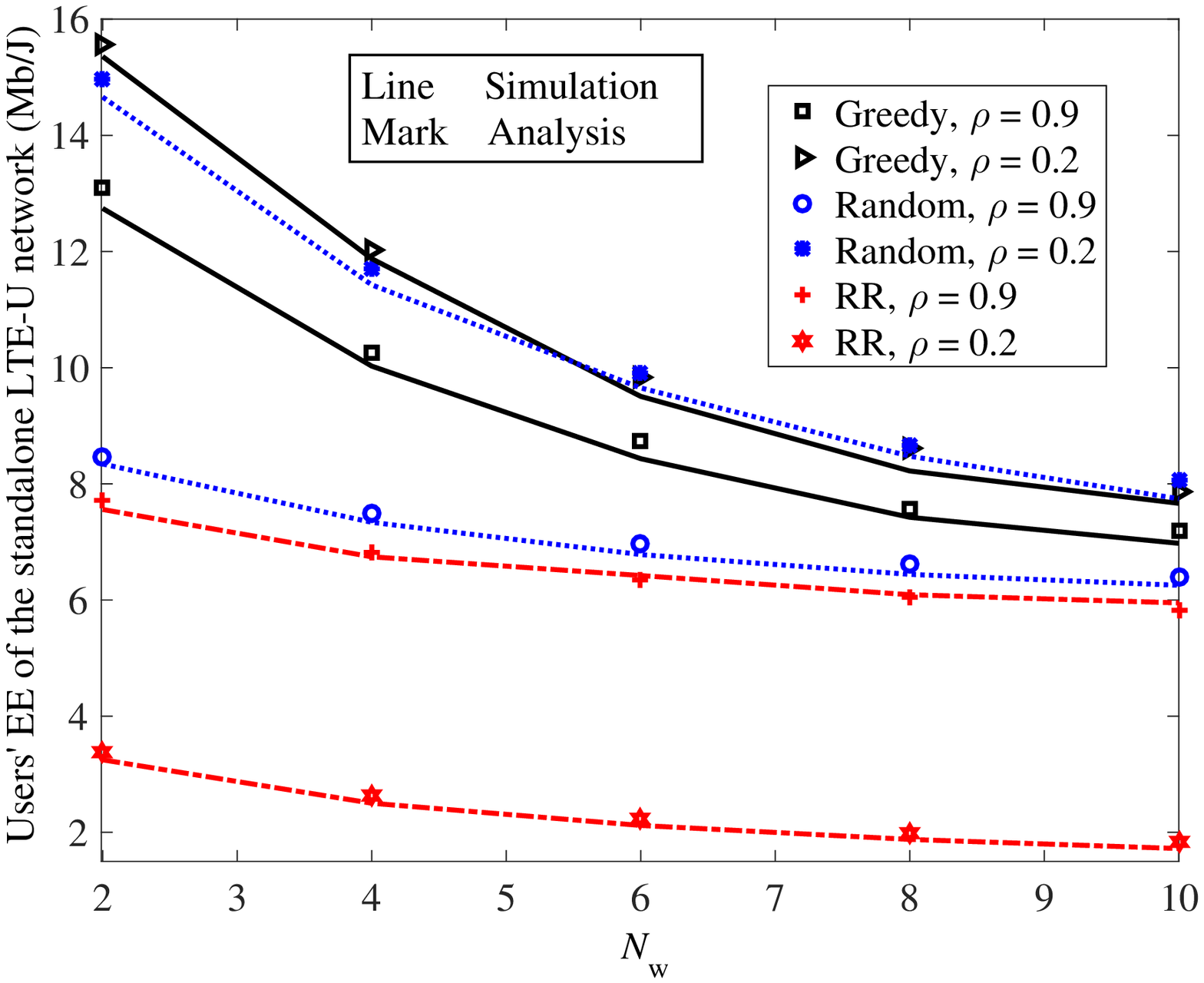}}
  \subfigure[]%
            {\label{fig:EE BM IID}%
             \includegraphics[width=0.49\textwidth]{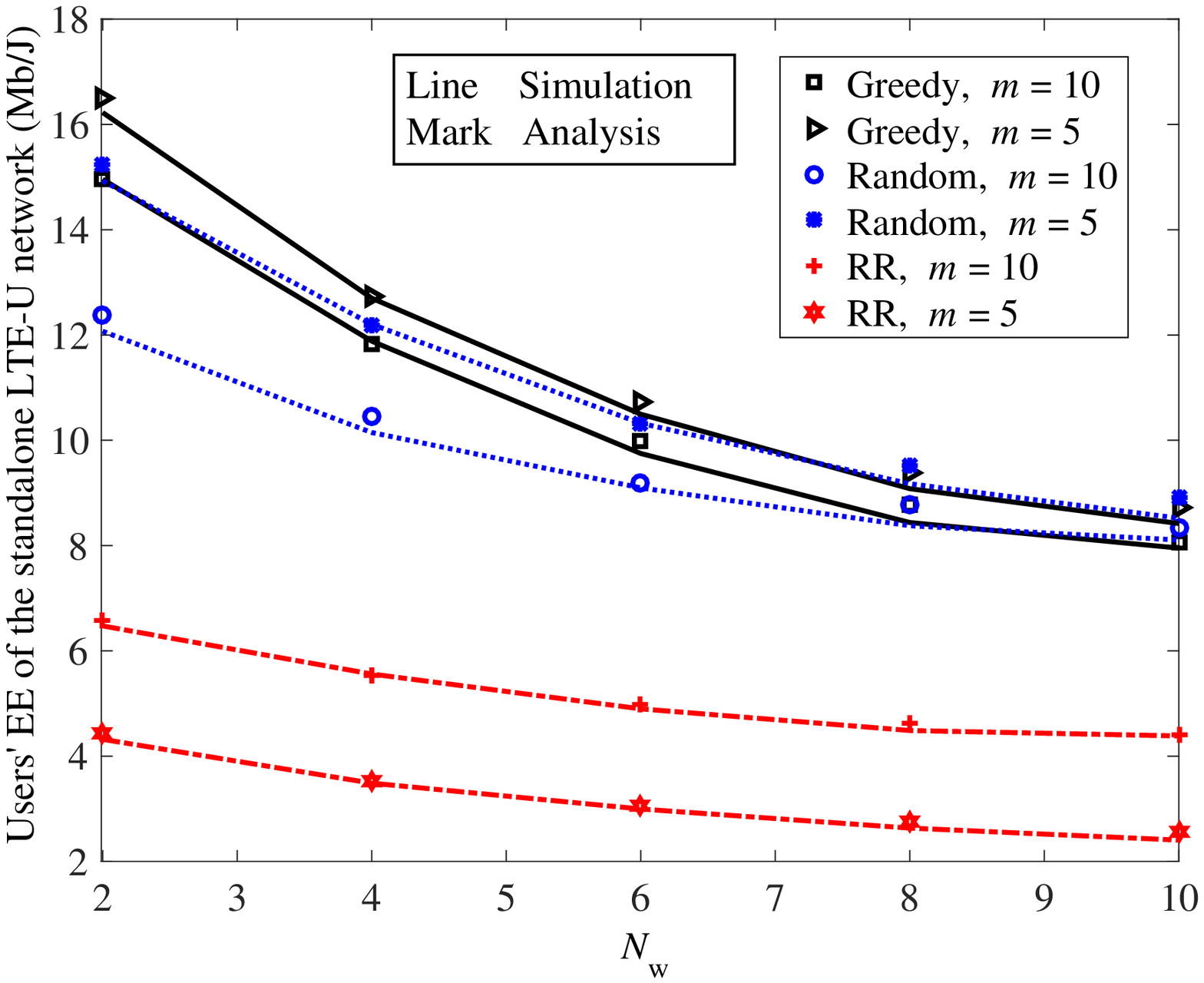}}
             \vspace{-0.15cm}
  \caption{Users' EE of the standalone LTE-U network in the i.i.d. user scenario. (a) EE under the threshold-based feedback scheme. (b) EE  under the best-$m$ feedback scheme.}
  \label{fig: EE IID}
  \vspace{-0.45cm}
\end{figure*}

In Fig. \ref{fig: EE IID}, we study users' EE of the standalone LTE-U network in the  i.i.d. user scenario.
It can be seen that for all schedulers under both the threshold-based feedback scheme and best-$m$ feedback scheme users' EE reduces as the Wi-Fi STA number increases.
This is because, as more Wi-Fi STAs contend channel, not only the throughput of the standalone LTE-U network reduces accordingly (see Figs. \ref{fig:througput TH} and \ref{fig:througput BM}),
but also the users consume more time on sensing channel, leading to both increased sensing energy consumption (see (\ref{eq: Expected energy consumption E u}) and (\ref{eq: Expected energy consumption E d})) and basic circuit energy consumption (see (\ref{eq: Expected energy consumption E ba})).
Further, we notice that for the greedy and random schedulers users' EE at $\rho = 0.2$ is higher than that at $\rho = 0.9$.
Taking the greedy scheduler as an example, as observed in Fig. \ref{fig:througput TH}, although the throughput of the scheduler at $\rho = 0.9$ is a little bit higher than that at $\rho = 0.2$, in the former case each user has more chances to not only report CSI of any specific subchannel but also be scheduled to receive data on that subchannel, i.e., consuming more energy on CSI feedback (see (\ref{eq: Expected E mk})) and data reception (see (\ref{eq: ECD Transmission})). 
That is, for the greedy scheduler, a higher throughput can be achieved under a greater $\rho$, but the EE performance is reduced in such a case.
%
In contrast, users' EE under the round robin scheduler at $\rho = 0.9$ is higher than that at $\rho = 0.2$.
This is because as compared with the increased energy consumption on CSI feedback and data reception, the network throughput here at $\rho = 0.2$ is much lower than that at $\rho = 0.9$.
It is also noteworthy that, for both feedback schemes, when the Wi-Fi STA number is larger than a certain value, users' EE under the random scheduler can be higher than that under the greedy scheduler, owing to the effectiveness of the random scheduler in dealing with long feedback delay (see Figs. \ref{fig:througput TH} and \ref{fig:througput BM}).

\begin{figure*}
  \centering
  \subfigure[]%
            {\label{fig:EE TH NIID}%
             \includegraphics[width=0.49\textwidth]{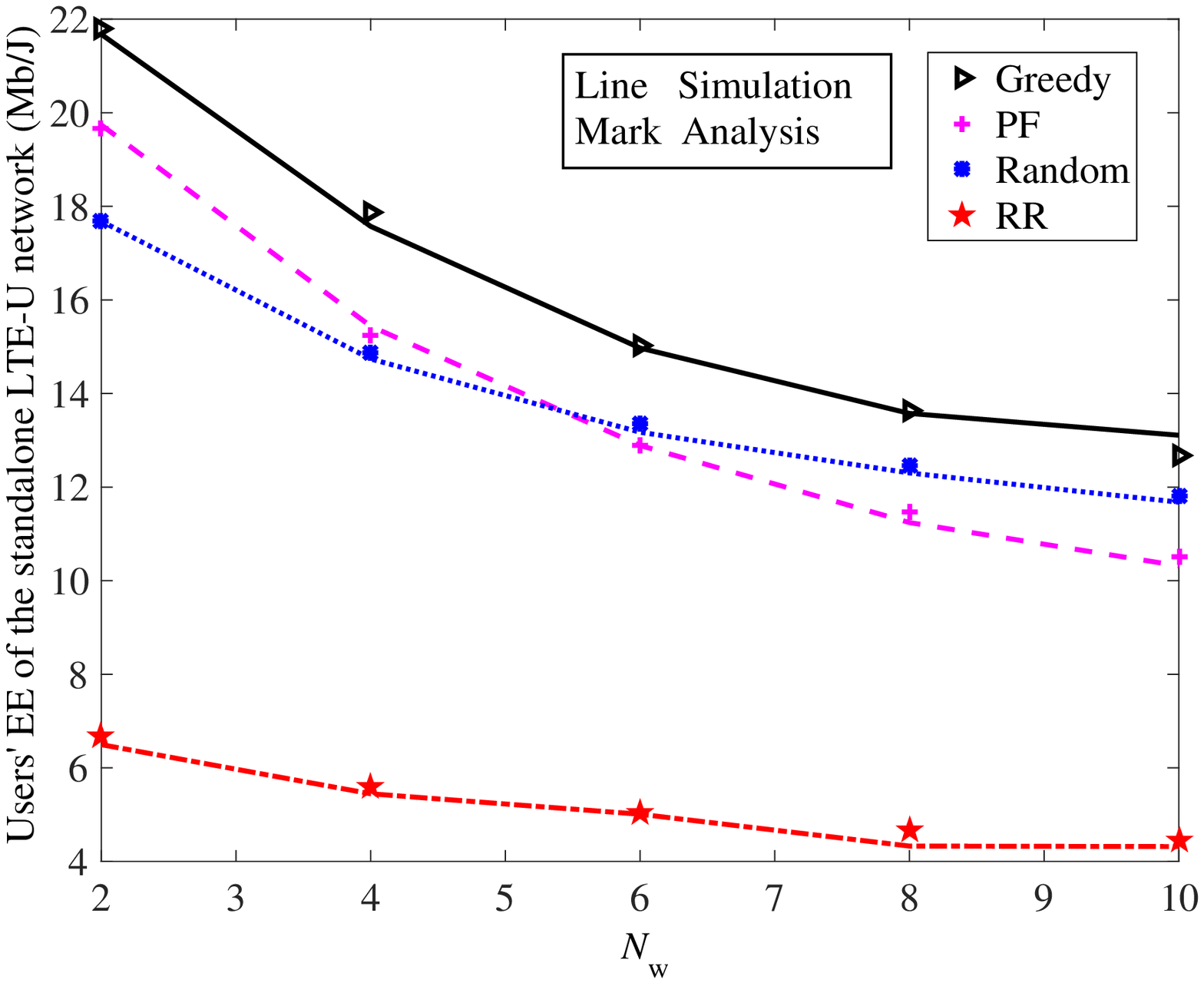}}
  \subfigure[]%
            {\label{fig:EE BM NIID}%
             \includegraphics[width=0.49\textwidth]{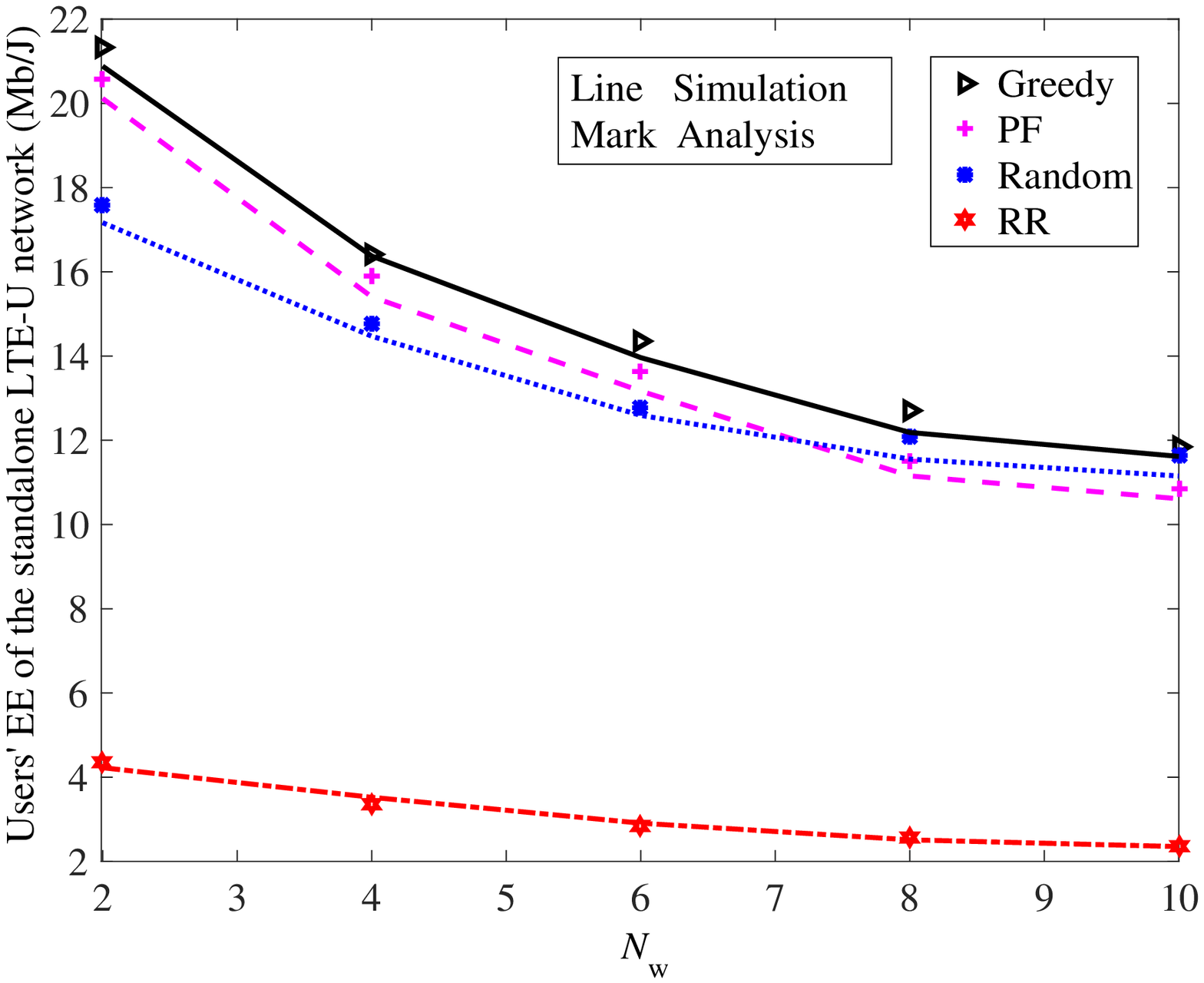}}
             \vspace{-0.15cm}
  \caption{Users' EE of the standalone LTE-U network in the non-i.i.d. user scenario. (a) EE under the threshold-based feedback scheme at $\rho = 0.5$. (b) EE under the best-$m$ feedback scheme at $m = 5$.}
  \label{fig: EE NIID}
  \vspace{-0.65cm}
\end{figure*}

For the non-i.i.d. user scenario, we study users' EE performance with the same setting for Section \ref{subsect: throughput noniid users}. As shown in Fig. \ref{fig: EE NIID}, we find that for a fixed $\rho$ (resp. $m$) of the threshold-based (resp. best-$m$) feedback scheme users' EE of the standalone LTE-U network changes with similar trends on throughput as shown in Fig. \ref{fig:throughput NIID}. However, if normalized by the performance of the round robin scheduler, we find that the mean EE improvements of the greedy, random, and PF schedulers are respectively lower than their mean network throughput improvements.
Taking the threshold-based feedback scheme as an example, the former are 214.5\%, 172.1\%, and 168.7\%, respectively, while the later are 221.6\%, 180.8\%, and 176.0\%, respectively.


\begin{figure}
  \centering
  \includegraphics[width=0.55\linewidth]{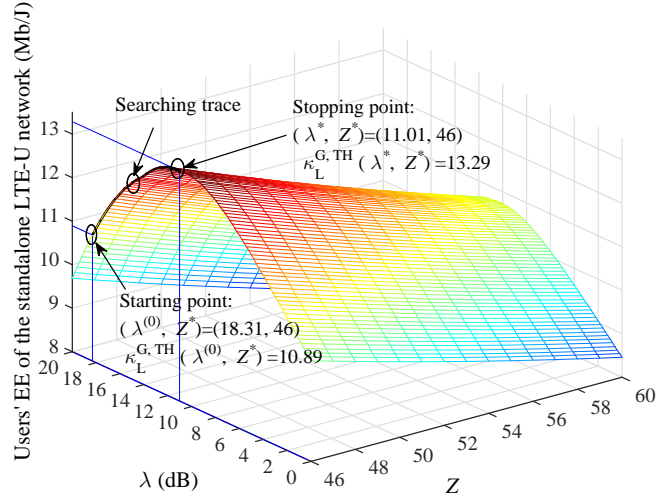}
  \vspace{-0.4cm}
  \caption{Searching trace of the optimized parameter setting of ($\lambda, Z$) for the threshold-based feedback scheme.}
  \label{fig:covergece TH}
  \vspace{-0.75cm}
\end{figure}

\subsection{EE Optimization}

In below we study the performance of the proposed coexistence-aware EE optimization algorithms.
For space limitation, only the results of the i.i.d. user scenario are presented.
Here, we assume that Wi-Fi's minimal channel occupancy time ratio threshold is set according to node numbers of the two coexisting networks, i.e., $D_{\rm th} = N_{\rm w}/(K+N_{\rm w})$. 
For the parameters in Algorithm 1, we set the stopping criterion $\beta =10^{-3}$ and the maximum threshold $\lambda_{\rm max} = 20$ dB. Other parameters are the same as those given in Section \ref{subsect: Parameter Settings}.


Fig. \ref{fig:covergece TH} shows an example of the searching trace of parameters $(\lambda, Z)$ for the threshold-based feedback scheme with the proposed algorithm, at $J=1$ and $N_{\rm w} = 6$ (thus $D_{\rm th} = 6/16 = 37.5\%$).
Specifically, the desired CW size $Z^* = 46$ is obtained by calculating $Z_0$; but, the desired feedback threshold $\lambda ^* =11.01$ is searched via Algorithm 1.
It can be observed that, for the tested scenario and given any fixed $\lambda$, users' EE reduces with $Z$, showing the effectiveness of the proposed approach in choosing CW size for the standalone LTE-U network.
Besides, numerical results unearth that with the optimized parameter setting $(\lambda ^*, Z^*)$ Wi-Fi's channel occupancy time ratio $t^{\rm w}_{\rm s}$ reaches 37.9\%, larger than the required $D_{\rm th}$.
It is also noteworthy from the figure that as compared with randomly setting $(\lambda, Z)$ from $[0, 20]$dB $\times [46, 60]$, users' EE with $(\lambda ^*, Z^*)$ increases 26.1\%.

\begin{table*}[!htbp]
\centering
\caption{Comparison between exhaustive search and the proposed algorithm for the threshold-based feedback scheme under different numbers of Wi-Fi STAs and/or node moving speeds.}
\vspace{-0.35cm}
\begin{tabular}{|c|c|c|c|c|c|}
\hline
\multicolumn{2}{|c|}{ \multirow{3}*{$(\lambda^*, Z^*, {\kappa _{\rm{L}}^{\rm G, TH}}^*)$} }& \multicolumn{4}{c|}{Moving speed} \\
\cline{3-6}
\multicolumn{2}{|c|}{ } & \multicolumn{2}{|c|}{2.0 (km/h) } & \multicolumn{2}{|c|}{ 3.0 (km/h) }\\
\cline{3-6}
\multicolumn{2}{|c|}{ } & Exhaustive search & Proposed algorithm & Exhaustive search & Proposed algorithm \\
\hline
\multirow{5}*{$N_{\rm w}$}
 & 2 & (12.40, 34, 26.54) & (11.97, 34, 26.51) & (12.20, 34, 21.08) & (12.01, 34, 20.61)  \\
\cline{2-6}
 & 4 & (12.20, 40, 22.99) & (11.76, 40, 22.57) & (11.80, 40, 17.46) & (11.83, 40, 17.03)  \\
\cline{2-6}
 & 6 & (12.00, 46, 19.81) & (11.53, 46, 19.37) & (11.40, 46, 13.76) & (11.01, 46, 13.29)  \\
\cline{2-6}
 & 8 & (11.60, 54, 16.75) & (10.50, 54, 16.32) & (10.40, 54, 11.54) & (9.72, 54, 11.17)  \\
\cline{2-6}
 & 10 &(10.60, 64, 13.81) & (10.20, 64, 13.45) & (10.00, 64, 9.03)  & (10.21, 64, 8.61)  \\
\hline
\end{tabular}
\label{tab: brute searching}
\vspace{-0.2cm}
\end{table*}

To understand the performance limit of the proposed algorithm for the threshold-based feedback scheme, more extensive simulations are run to compare the final parameter setting and the optimized users' EE between the proposed algorithm and exhaustive search.
As shown in Table \ref{tab: brute searching}, the proposed algorithm always finds the same CW size with exhaustive search but a local optimal value for the feedback threshold that however is close to the one found by exhaustive search.
Here, for computation time in exhaustive search the search step to identify the best $\lambda$ and the number of new searches $J$ in Algorithm 1 are set equal to 0.2 dB and 3, respectively.
Observed from simulation the main reason for the existence of a small mismatch between $\lambda$'s found by both algorithms is the non-convex and non-concave behavior in a certain neighborhood of the global optimal value.
However, if normalized by the optimal EE obtained by exhaustive search, the mean performance degradation is only 2.5\%.
In addition, from Table \ref{tab: brute searching} the following  observations are obtained.
Firstly, users' EE of the standalone LTE-U network reduces as users' moving speed increases. The performance decrease between moving speed equal to 2 km/h and 3 km/h can increase from 20.57\% to 34.61\%, as the number of Wi-Fi STAs increases from 2 to 10.
Secondly, to cope with the increase of Wi-Fi STAs (thus the increased difficulty to guarantee Wi-Fi's minimal channel occupancy time ratio) but improve users' EE in the LTE-U network as much as possible, CW size of the standalone LTE-U network should increase, while its channel feedback threshold should reduce thus gaining more opportunities to utilize channel, even it is poor.

\begin{figure}
  \centering
  \includegraphics[width=0.55\linewidth]{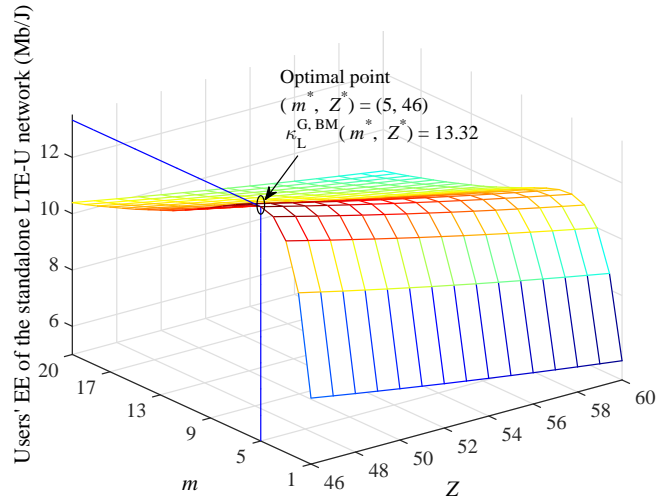}
  \vspace{-0.3cm}
  \caption{Searching trace of the optimized parameter setting of ($m, Z$) for the best-$m$ feedback scheme.}
  \label{fig:covergece BM}
  \vspace{-0.75cm}
\end{figure}

Fig. \ref{fig:covergece BM} shows an example of the searching trace of parameters $(m, Z)$ for the best-$m$ feedback scheme with the proposed algorithm at $N_{\rm w} = 6$.
As the CW size is obtained with the same approach for the threshold-based feedback scheme, here $Z^* = 46$ as well.
The best $m$ equal to 5 is found by exhaustive search in its limited value range.
It is found from simulation that for the tested scenario as compared with randomly setting $(m, Z)$ from $[1, 20] \times [46, 60]$, the increase of users' EE for the standalone LTE-U network with $(m^*, Z^*)$ can be as large as 27.1\%.



\section{Conclusion}
\label{sect: conclusion}


In this work, we have studied the impact of randomly delayed CSI on the performance of the standalone LTE-U network, by analyzing two performance metrics, i.e., the DL  throughput and users' EE, under two feedback schemes 
and four frequency-domain schedulers. 
Based on the analysis, 
we have further studied cross-layer coexistence-aware protocol parameter optimization for maximizing users' EE of the standalone LTE-U network.
Simulation results not only verify the accuracy of the analysis but also shed some light on designing a robust scheduler to deal with long feedback delay and setting both the MAC and physical layers' parameters to improve the users' EE of a standalone LTE-U network while protecting Wi-Fi.
Further works will be carried out to address feedback user selection design for the standalone LTE-U network.

\begin{appendices}

\section{Deduction of $\tau_{\rm w}$ and $p_{\rm L}$}
\label{app:deduce tau pL}

Let $\tau_{\rm w}$ and $\tau_{\rm L}$ denote the probability of a Wi-Fi STA sending a packet and that of the LTE-U BS or the users sending reservation signals in any backoff slot, respectively.
According to \cite{Bianchi2000Performance}, they can be obtained as
\begin{align} \label{eq: tau w}
{\tau _{\rm w}} = \frac{{2\left( {1 - 2{p_{\rm w}}} \right)}}{{\left( {1 - 2{p_{\rm w}}} \right)\left( {W + 1} \right) + {p_{\rm w}}W[ {1 - {{\left( {2{p_{\rm w}}} \right)}^{b_{\rm w}}}} ]}}
\end{align}%
\begin{align} \label{eq: tau L}
{\tau _{\rm L}} = 2/(Z+1) 
\end{align}
where $W$ and ${b_{\rm w}}$ are the minimum CW size and the backoff stage of the Wi-Fi network, respectively, and $Z$ is the fixed CW size of the LTE-U network.

Similar to $p_{\rm L}$, 
let $p_{\rm w}$ represent the probability of a Wi-Fi STA collided with other nodes in the network. It is obvious that $p_{\rm L}$ and $p_{\rm w}$ satisfy
\begin{align} \label{eq: pw}
{p_{\rm w}} = 1 - {\left( {1 - {\tau _{\rm w}}} \right)^{{N_{\rm w}} - 1}}\left( {1 - {\tau _{\rm L}}} \right)
\end{align}
\begin{align} \label{eq:  pL}
{p_{\rm L}} = 1 - {\left( {1 - {\tau _{\rm w}}} \right)^{{N_{\rm w}}}}
\end{align}
as a Wi-Fi STA accesses channel successfully only if neither other Wi-Fi SATs nor LTE-U BS or users access channel simultaneously, and the LTE-U BS or users access channel successfully only if all Wi-Fi SATs keep silently at the same time.
Solving (\ref{eq: tau w})-(\ref{eq:  pL}), we obtain not only $p_{\rm L}$ but also  $\tau_{\rm w}$, $\tau_{\rm L}$, and $p_{\rm w}$.

\vspace{-0.5cm}

\section{Proof of Theorem \ref{theo:1}}
\label{app:theo:1}

For the  i.i.d. user scenario, when the random scheduler and the threshold-based feedback scheme are adopted, given feedback delay $\tau _\alpha$, the mean transmission rate in the $s$-th subchannel at the $\alpha$-th subframe during a DL data transmission can be defined as
\begin{align} \label{eq: subchannel throughput of R TH iid}
\begin{split}
{\mathbb E} (\eta _{s,\alpha}^{\rm R,TH}| \tau _\alpha) = &  \sum\limits_{\delta  = 1}^K P\left\{ |{\cal U}_s^{\rm TH}| =  \delta |\tau _\alpha  \right\} 
\sum\limits_{n = \omega }^N  {r_n}  {P \left\{ { {G_{k_{\rm s},s,\alpha}^{\rm d}} \ge {L_n}, G_{k_{\rm s},s} \in \left[{\max}\left\{ {L_n,\lambda} \right\}, L_{n + 1} \right)    \left| |{\cal U}_s ^{\rm TH}| = \delta, \tau _\alpha \right. } \right\}}\\
=& \sum\limits_{\delta  = 1}^K \underbrace{P\left\{ |{\cal U}_s ^{\rm TH}| =  \delta  \right\}}_{(\rm a)} 
\sum\limits_{n = \omega }^N  {r_n}  \underbrace{P \left\{ { {G_{k_{\rm s},s,\alpha}^{\rm d}} \ge {L_n}, G_{k_{\rm s},s} \in \left[{\max}\left\{ {L_n,\lambda} \right\}, L_{n + 1} \right) \left| |{\cal U}_s ^{\rm TH}| = \delta, \tau _\alpha \right. }  \right\}}_{(\rm b )}
\end{split}
\end{align}
where $\omega$ is the index such that $L_{\omega} \le \lambda < L_{\omega + 1}$ holds, ${\cal U}_s ^{\rm TH}$ is the user set in which all users fed back their channel gains for subchannel $s$ according to the threshold-based feedback scheme in the previous UL transmission, and $k_{\rm s}$ is the index of user scheduled for data reception in subchannel $s$ at subframe $\alpha$. In general, the size of ${\cal U}_s ^{\rm TH}$, $\delta$, is between 0 and $K$; yet, only if the set is not empty (i.e., $\delta \geq 1$), the subchannel can be assigned to a user for data transmission and contributes non-zero data rate.
The second equality holds as the two event $|{\cal U}_s ^{\rm TH}| =  \delta$ and feedback delay equal to $\tau _\alpha$ are independent.

For probability (a) in (\ref{eq: subchannel throughput of R TH iid}), in the i.i.d. user scenario since the subchannel gains of different users are statistically identical and independent, it can be calculated as
\begin{align} \label{eq: prob in set U G TH iid}
\begin{split}
P\left\{ |{\cal U}_s ^{\rm TH}| =  \delta  \right\} 
= \binom{K}{\delta}{\left( {\int_\lambda ^\infty  {f_G(g) dg} } \right)^\delta }{\left( {\int_0^\lambda  { f_G(g) dg} } \right)^{K - \delta }} 
=  \binom{K}{\delta}{\exp}\left( { - \frac{{\lambda \delta }}{\Omega }} \right){\left( {1 - {\exp}\left( { - \frac{\lambda }{\Omega }} \right)} \right)^{K - \delta }}
\end{split}
\end{align}
where $f_G(g) = (1/\Omega) {\exp}\left({ - {{\rm{g}}}/{\Omega}} \right)$ is the power distribution of the Rayleigh fading channel for any user \cite{goldsmith2005wireless}.
For probability (b) in (\ref{eq: subchannel throughput of R TH iid}), we can figure it out as follows
\begin{align} \label{eq: prob R TH iid x set U tau}
\begin{split}
&P \left\{ {G_{k_{\rm s},s,\alpha}^{\rm d}} \ge {L_n},{\max}\left\{ {{L_n},\lambda } \right\} \le {G_{k_{\rm s},s}} < {L_{n + 1}}\left| |{\cal U}_s ^{\rm TH}| = \delta,{\tau _\alpha } \right. \right\} \\
& = \binom{\delta}{1} \frac{1}{\delta} P\left\{ {G_{1,s,\alpha}^{\rm d}} \ge {L_n},{\max}\left\{ {{L_n},\lambda } \right\} \le {G_{1,s}} < {L_{n + 1}} \left| {\cal U}_s ^{\rm TH} = \{{\rm users}~1,2,...,\delta\}, {\tau _\alpha } \right. \right\} \\
&= P\left\{ {G_{1,s,\alpha}^{\rm d}} \ge {L_n},{\max}\left\{ {{L_n},\lambda } \right\} \le {G_{1,s}} < {L_{n + 1}} \left|G_{1,s} \ge \lambda, {\tau _\alpha } \right. \right\} \\
&= 
{ P \left\{ {G_{1,s,\alpha}^{\rm d}} \ge {L_n},{\max}\left\{ {{L_n},\lambda } \right\} \le {G_{1,s}} < {L_{n + 1}}, G_{1,s} \ge \lambda \left| {\tau _\alpha } \right. \right\}}/{P\{G_{1,s} \ge \lambda \}} \\
&= \exp{\left(\frac{\lambda}{\Omega}\right)} P \left\{ {G_{1,s,\alpha}^{\rm d}} \ge {L_n},{\max}\left\{ L_n,\lambda  \right\} \le {G_{1,s}} < {L_{n + 1}}, G_{1,s} \ge \lambda \left| {\tau _\alpha } \right. \right\} \\
&=\exp{\left(\frac{\lambda}{\Omega}\right)} \int_{{\max}\left\{ {{L_n},\lambda } \right\}}^{{L_{n + 1}}}  \int_{{L_n}}^\infty  f_{{G_{1,s}},{G_{1,s,\alpha}^{\rm d}} }\left( {x,y|{\tau _\alpha }} \right)dydx
\end{split}
\end{align}
where the first equation holds because one of $\delta$ users should be selected from ${\cal U}_s ^{\rm TH}$ and each of them is selected with equal probability $1/\delta$;
the second equation holds due to the independence of the channel gains of subchannel $s$ of different users;
the forth equation holds because $P\{G_{1,s} \geq \lambda\} =\int\nolimits_{\lambda}^{\infty}f_G(g) dg $.
%
%
Substituting (\ref{eq: joint pdf}), (\ref{eq: prob in set U G TH iid}), and (\ref{eq: prob R TH iid x set U tau}) into (\ref{eq: subchannel throughput of R TH iid}), we end the proof of Theorem \ref{theo:1}.

\section{Proof of Theorem \ref{theo:2}}
\label{app:theo:2}

Similar to Appendix \ref{app:theo:1}, for the i.i.d. user scenario, when the random scheduler and the best-$m$ feedback scheme are adopted, given feedback delay $\tau _\alpha$, the mean transmission rate in the $s$-th subchannel at the $\alpha$-th subframe during a DL data transmission can be formulated as
\begin{align} \label{eq: subchannel throughput of R BM iid}
\begin{split}
{\mathbb E} (\eta _{s,\alpha }^{\rm R,BM}| \tau _\alpha) =&   \sum\limits_{\delta  = 1}^K \underbrace{P\left\{ |{\cal U}_s^{\rm BM}| =  \delta  \right\}}_{(\rm a)} \sum\limits_{n = 1}^N  {r_n}
 \underbrace{P\left\{ { {G_{k_{\rm s}, s,\alpha}^{\rm d}} \ge {L_n}, {G_{k_{\rm s},s}} \in \left[ {L_n}, {L_{n + 1}}\right) \left| |{\cal U}_s^{\rm BM}| = \delta, \tau _\alpha  \right.} \right\}}_{(\rm b)}
\end{split}
\end{align}
where ${\cal U}_s^{\rm BM}$ is the user set in which all users fed back their channel gains for subchannel $s$ according to the best-$m$ feedback scheme in the previous UL transmission.
Different from  (\ref{eq: subchannel throughput of R TH iid}), in (\ref{eq: subchannel throughput of R BM iid}) data rate can span from $r_1$ rather than $r_ \omega$ to $r_N$, as it is possible that the channel quality of the best channel fed back only offers the lowest data rate.
For probability (a) in (\ref{eq: subchannel throughput of R BM iid}), it can be obtained as
\begin{align} \label{eq: prob in set U G BM iid}
P\left\{ |{\cal U}_s^{\rm BM}| =  \delta  \right\} =  \binom{K}{\delta}{\left( {\frac{m}{S}} \right)^\delta }{\left( {1 - \frac{m}{S}} \right)^{K - \delta}}
\end{align}
where $m/S$ gives the probability of each user feeding back CSI for subchannel $s$ according to the best-$m$ feedback scheme  in the i.i.d. user scenario.
Here, for probability (b), let ${\Xi _{k,s}}$ be the number of user $k$'s subchannels of which channel quality as estimated is better than  subchannel $s$. So, if $G_{k,s}$ is within the best $m$ subchannel gains, ${\Xi _{k,s}} \le m - 1$. Then, we can derive probability (b) as follows
\begin{align} \label{eq: prob R BM iid x set U tau}
\begin{split}
& P \left\{ {G_{k_{\rm s},s,\alpha}^{\rm d}} \ge {L_n}, {L_n} \le {G_{k_{\rm s},s}} < {L_{n + 1}}\left| |{\cal U}_s^{\rm BM}| = \delta,{\tau _\alpha } \right. \right\} \\
&= \binom{\delta}{1} \frac{1}{\delta}  P \left\{ {G_{1,s,\alpha}^{\rm d}} \ge {L_n}, {L_n}  \le {G_{1,s}} < {L_{n + 1}}, \left|{\cal U}_s^{\rm BM} = \{{\rm users}~1,2,...,\delta\} , {\tau _\alpha } \right. \right\} \\
&=  P \left\{ {G_{1,s,\alpha}^{\rm d}} \ge {L_n},  {L_n} \le {G_{1,s}} < {L_{n + 1}}, \left| {\Xi _{1,s}} \le {m - 1}, {\tau _\alpha } \right. \right\} \\
&= 
{ P \left\{ {G_{1,s,\alpha}^{\rm d}} \ge {L_n},  L_n  \le {G_{1,s}} < {L_{n + 1}}, {\Xi _{1,s}} \le {m - 1} \left| {\tau _\alpha } \right. \right\}}/{P\{{\Xi _{1,s}} \le {m - 1}\}} \\
&=  \frac{S}{m} \cdot  P \left\{ {G_{1,s,\alpha}^{\rm d}} \ge {L_n}, L_n \le {G_{1,s}} < {L_{n + 1}}, {\Xi _{1,s}} \le {m - 1} \left| {\tau _\alpha } \right. \right\} \\
&=  \frac{S}{m} \cdot P \left\{{\Xi _{1,s}} \le {m - 1}| {G_{1,s,\alpha}^{\rm d}} \ge {L_n}, {G_{1,s}} \in \left[L_n, L_{n + 1}\right), {\tau _\alpha} \right\} P \left\{{G_{1,s,\alpha}^{\rm d}} \ge {L_n}, {G_{1,s}} \in \left[L_n, L_{n + 1}\right) | {\tau _\alpha} \right\} \\
&= \frac{S}{m} \cdot P \left\{{\Xi _{1,s}} \le {m - 1}| L_n \le {G_{1,s}} < {L_{n + 1}}\right)  P \left({G_{1,s,\alpha}^{\rm d}} \ge {L_n}, L_n \le {G_{1,s}} < {L_{n + 1}} | {\tau _\alpha} \right\} \\
&= \frac{S}{m} \cdot \int_{L_n}^{{L_{n + 1}}}  \int_{{L_n}}^\infty  P \{{\Xi _{1,s}} \le {m - 1}\left| G_{1,s} = x \right.\} f_{{G_{1,s}},{G_{1,s,\alpha}^{\rm d}} } \left( {x,y|{\tau _\alpha }} \right)dydx
\end{split}
\end{align}
where the first two equations hold due to the same reasons for the first two equations of (\ref{eq: prob R TH iid x set U tau});
the forth equation holds as $P\{{\Xi _{1,s}} \le {m - 1}\} = m/S$;
for the sixth equation we use the independent relation among the involved events.
For $P ({\Xi _{1,s}} \le {m - 1}\left| G_{1,s} = x \right.)$ in (\ref{eq: prob R BM iid x set U tau}), it has been derived in \cite{Guharoy2013Joint} as
\begin{align} \label{eq: prob Gks x}
P \left\{ {{\Xi _{1,s}} \le m - 1\left| {{G_{1,s}} = x} \right.} \right\}  = \sum\limits_{n = 0}^{m - 1} \binom{S-1}{n} {\rm exp}\left(  - \frac{xn}{\Omega } \right) \left( 1 - {\rm exp} \left(  - \frac{x}{\Omega} \right) \right)^{S-1-n}.
\end{align}
Finally, substituting (\ref{eq: joint pdf}) and (\ref{eq: prob in set U G BM iid}) - (\ref{eq: prob Gks x}) into (\ref{eq: subchannel throughput of R BM iid}), we finish the proof of Theorem \ref{theo:2}.

\end{appendices}

%
%
%
%
%

{\small
\bibliographystyle{IEEEtran}
\bibliography{PassiveCom}
}

\end{document}